\documentclass[11pt]{article}
\usepackage{graphicx}
\usepackage{amssymb}
\usepackage{amscd}
\usepackage{mathrsfs}
\usepackage{longtable,lscape}
\usepackage{amsthm}
\usepackage{amsfonts}
\usepackage{amsmath}
\usepackage{bbm}
\usepackage{float}
\usepackage{url}
\usepackage{hyperref}
\usepackage[margin=0.5cm,font=footnotesize]{caption}
\usepackage{lineno}
\usepackage[round]{natbib}
\usepackage{appendix}
\usepackage{subfig}

\begin{document}
\title{Are Words the Quanta of Human Language? Extending \\ the Domain of Quantum Cognition}
\author{Diederik Aerts and Lester Beltran  \vspace{0.5 cm} \\ 
        \normalsize\itshape
        Center Leo Apostel for Interdisciplinary Studies
        \\ 
        \normalsize\itshape
         Brussels Free University, Krijgskundestraat 33 \\ 
        \normalsize\itshape
         1160 Brussels, Belgium \\
        \normalsize
        E-Mails: \url{diraerts@vub.ac.be, diraerts@gmail.com},
           \\ \url{lbeltran@vub.ac.be, lestercc21@yahoo.com}
              	\\
              }
\date{}
\maketitle
\begin{abstract}
\noindent
In previous research, we showed that `texts that tell a story' exhibit a statistical structure that is not Maxwell--Boltzmann but Bose--Einstein. Our explanation is that this is due to the presence of 'indistinguishability' in human language as a result of the same words in different parts of the story being indistinguishable from one another, in much the same way that 'indistinguishability' occurs in quantum mechanics, also there leading to the presence of Bose--Einstein rather than Maxwell--Boltzmann as a statistical structure. In the current article, we set out to provide an explanation for this Bose--Einstein statistics in human language. We show that it is the presence of `meaning' in `texts that tell a story' that gives rise to the lack of independence characteristic of Bose--Einstein, and provides conclusive evidence that `words can be considered the quanta of human language', structurally similar to how `photons are the quanta of electromagnetic radiation'. Using several studies on entanglement from our Brussels research group, we also show, by introducing the von Neumann entropy for human language, that it is also the presence of `meaning' in texts that makes the entropy of a total text smaller relative to the entropy of the words composing it. We explain how the new insights in this article fit in with the research domain called `quantum cognition', where quantum probability models and quantum vector spaces are used in human cognition, and are also relevant to the use of quantum structures in information retrieval and natural language processing, and how they introduce `quantization' and `Bose--Einstein statistics' as relevant quantum effects there. Inspired by the conceptuality interpretation of quantum mechanics, and relying on the new insights, we put forward hypotheses about the nature of physical reality. In doing so, we note how this new type of decrease in entropy, and its explanation, may be important for the development of quantum thermodynamics. We likewise note how it can also give rise to an original explanatory picture of the nature of physical reality on the surface of planet Earth, in which human culture emerges as a reinforcing continuation of life.
\end{abstract}

\section{Introduction \label{introduction}}
If we consider `a text that tells a story', then the same words can be interchanged within the text without changing anything about the story told by the text. Let us demonstrate what we mean by using a concrete text which we call the `Example Text'  

\begin{quotation}
\noindent
It mists over the flanks of the hillside. In a few hours the climbing sun will dispel the haze. The pine trees are slender against the horizon. The \underline{animal} saunters slowly down toward the great river. Before reaching the water, the \underline{animal} looks into the distance and lets its proud \underline{sound} reverberate across the valley. The \underline{animal} continues its way through the young grass to the riverbank where it will enjoy its favorite \underline{food}. From the small grove of trees growing a little further along the riverbank, another \underline{animal} appears, surveying the panorama.
\end{quotation}

\noindent
We have underlined the four occurrences of the word {\it{Animal}} 
 in the `Example Text', and it is indeed evident that we can interchange these four words {\it Animal} with each other, and that this will not change the 
text as a story in any possible way (we will denote `words' in italics and with an initial capital letter like we have denoted `concepts' in our earlier works). Herein, `a text representing a story' behaves differently from `a collection of macroscopic physical objects representing a landscape' in physical reality. Indeed, let us now consider the physical landscape of `the animal sauntering down the hillside and the other animal appearing from the grove of trees along the riverbank'. If we swap these two animals in the physical landscape, this landscape will change essentially. 

Macroscopic material entities, such as animals, even if they are designated by the same word in a text, cannot be exchanged without fundamentally changing the physical reality of the landscape in which they are present. Indeed, two macroscopic physical objects, such as the one animal and the other animal in our example of the landscape of physical objects described by our `Example Text', are never identical and indistinguishable. On the other hand, two words {\it Animal} in the text of a story, designating the one {\it Animal} and the other {\it Animal} in our `Example Text', are definitely identical and indistinguishable. This insight inspired us to investigate the statistical nature of the distribution of words in the texts of stories, paying attention to this symmetry of identity and indistinguishability. We were able to show that this distribution can be modeled in an exceptionally accurate way by Bose--Einstein statistics, while Maxwell--Boltzmann statistics is totally incapable of providing a fitting model. We tested the two statistics, Bose--Einstein and Maxwell--Boltzmann, on many different texts representing stories, short stories, and long stories of the size of novels. We observed that time and again a Bose--Einstein distribution yielded a perfect fit, whereas the Maxwell--Boltzmann distribution never came even close to a fitting distribution \citep{aertsbeltran2020}. The reader may have noticed that in the `Example Text' we have also underlined the words {\it Sound} and {\it Food}. The reason is that these words will be used in Section \ref{entropyandpurity} to illustrate our new insight related to the presence of entanglement in human language.

We note that it is one of the basic principles of quantum mechanics that in a physical landscape of quantum particles the same particles are identical and indistinguishable. By expressing in the quantum formalism that particles are identical and indistinguishable, it can be shown that their wave functions are either symmetric or anti-symmetric. Additionally, it can be shown that, for the case of the symmetric wave functions, the statistics is Bose--Einstein, and we call such particles bosons, and that, for the case of the anti-symmetric wave functions, it is Fermi--Dirac, and we call such particles fermions. On the other hand, an equivalent mathematical derivation leads to the Maxwell--Boltzmann statistics when dealing with classical distinguishable particles. This means that identifying one of the three statistics, Bose--Einstein, Fermi--Dirac, or Maxwell--Boltzmann, is like finding a fingerprint that uniquely reflects what the particles `are', i.e. what their ontology is, namely either indistinguishable and bosonic, indistinguishable and fermionic, or distinguishable \citep{huang1987,yokoiabe2018}.

We wish in this article to continue working on our finding in \citet{aertsbeltran2020}, which was preceded by an initial examination in \citet{aertssozzoveloz2015b}, and which consists in considering `words as the quanta of human language'. Hence, the ground of our motivation is the perspective that `human language and its words' are similar to `ideal gases and their atoms or molecules' and also to `electromagnetic radiation and its photons'. 
With respect to the first comparison, that of human language with an ideal gas consisting of bosonic atoms or molecules, we already investigated aspects of this similarity in \citet{aertsbeltran2020}. More concretely, we compared the bosonic statistical structure we discovered in human language with the situation of a gas of bosonic atoms close to the Bose--Einstein condensate of this gas \citep{cornellwieman2002,ketterle2002}. 

In Section \ref{wordsasquanta}, we give one of the examples of a text for which we determined the Bose--Einstein distribution in \citet{aertsbeltran2020}. With respect to the second comparison, that of human language and its words with electromagnetic radiation and its photons, we will examine it in Section \ref{identifyindistinguishability}, using in part a historical account in which Max Planck and Albert Einstein in particular played a prominent role. This is also the way we put forward the notion of `quantization', originally proposed by Max Planck in the article that initiated what is often called the `Old Quantum Theory' \citep{planck1900}, subsequently further elaborated by Albert Einstein in the article which would ultimately give rise to his receiving the Nobel Prize, and where he uses Planck's quanta to describe the photoelectric effect \citep{einstein1905}.   

In this respect, our investigation also pertains to a research domain called quantum cognition \citep{aertsaerts1995,khrennikov1999,atmanspacher2002,busemeyeretal2006,aerts2009a,pothosbusemeyer2009,bruzaetal2009,lambertmoglianskizamirzwirn2009,busemeyerbruza2012,dallachiaraetal2012,havenkhrennikov2013,sozzo2014,aertssozzoveloz2015a,melucci2015,velozdujardin2015,blutnerbeimgraben2016,moreirawichert2016,broekaertetal2017}, where quantum probability models and quantum vector spaces are used to describe human decisions that depart from rationality (conjunction effect, disjunction effect, guppy effect, Ellsberg-type paradoxes, etc.). More specifically, we want to enlarge the domain of quantum cognition to the study of the quantum statistical properties of the cognitive apparatus that humans use in cognition, communication, and decision making, namely human language itself. More concretely, we want to introduce that aspect of quantum mechanics which is the oldest one, namely the aspect of `quantization', and more precisely consider `words' as the `quanta' of human language. In this way, we want to show that, in the research domain of quantum cognition, not only do quantum probability models and quantum vector spaces play a role, but quantization as well appears as a quantum effect. In Section \ref{identifyindistinguishability}, we also bring out the first of the two new insights in this article, namely an explanation for the presence of the Bose--Einstein statistics in human language. We show that the peculiar probabilities associated with Bose--Einstein statistics, leading to heated discussions among the protagonists of the Old Quantum Theory, are due to the presence of `meaning' in a story.

In Section \ref{entropyandpurity} we introduce the `von Neumann entropy' in human human language and show how it is greater than zero for single words while the entropy of the total text is zero, thus demonstrating the subadditivity of entropy in human language, very analogous to what takes place for quantum entities. To this end, we use the data from five studies of combinations of words that violate Bell's inequalities \citep{bell1964} from previous works by our Brussels research group over the last decade \citep{aertssozzo2011,aertsetal2021b,beltrangeriente2019,aertsarguelles2018}. This illustrates the second of our new insights, namely that in human language, a text telling a story possesses a lower entropy than each of its words. Additionally, our analysis shows that it is the entanglement of those words caused by the presence of `meaning' that is responsible for this decrease of entropy as a consequence of the composition of the text by its words.

We reflect on the impact of these two new insights in relation to different aspects of the nature of reality.

\section{Words as the Quanta of Human Language \label{wordsasquanta}}

We consider as an example again the first text we examined in \citet{aertsbeltran2020}, namely the Winnie the Pooh story entitled `In Which Piglet Meets a Haffalump' \citep{milne1926}.  

The `energy level' of a word is defined by the number of times this word appears in the considered text. The word that appears most often in the considered Winnie the Pooh story, namely 133 times, is the word {\it And} and we attribute to it the lowest energy level $E_0$. The second most-often appearing word, 111 times, is {\it He}, and we attribute to it the second-lowest energy level $E_1$, and so on, till we reach words such as {\it Able}, which appear only once.  

For a `gas of bosonic particles' in `thermal equilibrium with its environment', indeed these `numbers of times of appearance' indicate different energy levels of the particles of the gas, following the `energy distribution law governing the gas', and this is our inspiration for the introduction of `energy' in human language. We want to look at each of these words as `states of the quantum of human language', which we have called `the cogniton' in \citet{aertsbeltran2020}. This is in analogy with how the different energy levels of photons (different wavelengths of light) are each `states of the photon'. Proceeding in this way,
we arrive at 452 energy levels for the story `In Which Piglet Meets a Haffalump', the values of which are taken to be
\begin{eqnarray} \label{Ei}
 \{E_i = i \  |\ i \in [0, 1, \ldots, 451, 452]\}
 \end{eqnarray}

We denote $N(E_i)$ the `number of appearances' of the word with energy level $E_i$, and if we denote $n$ the total number of energy levels, we have that
\begin{eqnarray}
N = \sum_{i=0}^n N(E_i)
\end{eqnarray}
is the total number of words of the considered text, which is 2655 for the story `In Which Piglet Meets a Haffalump'.

For each of the energy levels $E_i$, $N(E_i)E_i$ is the amount of energy `radiated' by the story `In Which Piglet Meets a Haffalump' with the `frequency or wave length' connected to this energy level. For example, the energy level $E_{54} = 54$ is populated by the word {\it Thought} and the word {\it Thought} appears $N(E_{54})=10$ times in the story `In Which Piglet Meets a Haffalump'. Each of the 10 appearances of {\it Thought} radiates with energy value 54, which means that the total radiation with the wavelength connected to {\it Thought} of the story `In Which Piglet Meets a Haffalump' equals $N_{54}E_{54} = 10 \times 54 = 540$.

The total energy $E$ radiated by the considered text is therefore 
\begin{eqnarray}
E = \sum_{i=0}^n N(E_i) E_i
\end{eqnarray}

For the story `In Which Piglet Meets a Haffalump', we have $E = 242{,}891$. In Table \ref{piglethaffalunmp}, we have presented the list of all words that appear in the story (in the column `Words'), with their `number of appearances' (in the column `Appearance numbers $N(E_i)$'), ordered from lowest energy level to highest energy level (in the column `Energy levels $E_i$'), where the energy levels are attributed according to these numbers of appearances,  with lower energy levels attributed to a higher number of appearances, and their values are given as proposed in (\ref{Ei}).

Before we continue, however, we will briefly review the aspects of human language that are important to our view of it, which originated in the period when we were studying concepts of human language leading up to the emergence of quantum cognition \citep{gaboraaerts2002,aertsgabora2005a,aertsgabora2005b}. We consider a word or a concept as an entity that can be in different states and thus introduce our interpretation of what in concept research is called `prototype theory', originally introduced by Eleanor Rosch \citep{rosch1978,rosch1983}. We explicitly introduce the notion of `state of a concept', where this is done implicitly in the traditional version of prototype theory by means of the properties. The concept {\it Horse} is a state of the concept {\it Animal}, so an `exemplar' is a `state' of the concept of which it is an exemplar. However, that is not the only way in which states of concepts form. Likewise, when the word `animal' is in a text, the presence of the text surrounding the word `animal' changes the state of the corresponding concept {\it Animal}. Exemplars of a concept are states of that concept, but also `contexts that surround a concept in a text' determine states of that concept. Originally we introduced this view of concepts and words with the intention of describing and explaining effects such as `the guppy effect', and we used the formalism of quantum mechanics in a complex Hilbert space to do so \citep{gaboraaerts2002,aertsgabora2005a,aertsgabora2005b}. Note in that sense that in \citet{aertsbeltran2020} we introduced the new notion of `cogniton', to denote a `quantum of human language' and one can see that in Table \ref{piglethaffalunmp}, in the first column, we mention the three, `words', `concepts', and `cognitons'. In Section \ref{entropyandpurity}, we will explain still more in detail how our notion of `word' and `concept' or `cogniton', answers other questions one might put forward.  

The question can be asked, `what is the unity of energy in this model?', is the number `1' that we choose for energy level $E_1$ a quantity expressed in joules, or in electronvolts, or still in another unity? This question gives us the opportunity to reveal already one of the very new aspects of our approach. Energy will not be expressed in `${\rm kg m}^2/{\rm s}^2$', like it is the case in physics. Why not? Well, a human language is not situated somewhere in space, like we believe it to be the case with a physical boson gas of atoms, or a photon gas of light. Hence, in our approach here, `energy' is a basic quantity, and if we manage to introduce what the `human language equivalent' of `physical space' is---this is one of our aims in further work---then it will be oppositely, namely this `equivalent of space' will be expressed in unities where `energy appears as a fundamental unit'. Hence, the `1' indicating that `{\it He} radiates with energy 1', or `the cogniton in state {\it He} carries energy 1', stands with a basic measure of energy, just like `distance (length)' is a basic measure in `the physics of space and objects inside space', not to be expressed as a combination of other physical quantities. We used the expressions `{\it He} radiates with energy 1', and `the cogniton in state {\it He} carries energy 1', and we will use this way of speaking about `human language within the view of a boson gas of entangled cognitons that we develop here', in similarity with how we speak in physics about light and photons. 

The words {\it The}, {\it It}, {\it A}, and {\it To} are the four next most often appearing words of the Winnie the Pooh story, and hence the energy levels $E_2$, $E_3$, $E_4$, and $E_5$ are populated by cognitons respectively in the states {\it The}, {\it It}, {\it A}, and {\it To} carrying respectively 2, 3, 4, and 5~basic energy units.
Hence, the first three columns in Table \ref{piglethaffalunmp} describe the experimental data that we extracted from the Winnie the Pooh story `In Which Piglet Meets a Haffalump'. As we said, the story contains a total of 2655 words, which give rise to 542 energy levels, where energy levels are connected with words. Hence different words radiate with different energies, and the sizes of the energies are determined by `the number of appearances of the words in the story'. The most often appearing words are states of lowest energy of the cogniton and the least often appearing words are states of highest energy of the cogniton. In Table \ref{piglethaffalunmp}, we have not presented all 542 energy levels, because that would make the table impractically long, but we have presented the most important part of the energy spectrum with respect to the further aspects we will point out. 

More concretely, we have represented the range from energy level $E_0$, the ground state of the cogniton, which is the cogniton in state {\it And}, to energy level $E_{78}$, which is the cogniton in state {\it Put}. Then we have represented the energy level from $E_{538}$, which is the cogniton in state {\it Whishing}, to the highest energy level $E_{542}$ of the Winnie the Pooh story, which is the cogniton in state {\it You've}. The set of omitted words, between {\it Put} and {\it Wishing} are indicated by a blank space in Table \ref{piglethaffalunmp}.

These last five highest energy levels, from $E_{538}$ to $E_{542}$, corresponding respectively to the cogniton in states {\it Whishing}, {\it Word}, {\it Worse}, {\it Year}, and {\it You've}, words that all appear just once in the story. They do however radiate with different energies, but the story is not giving us enough information to determine whether {\it Whishing} is radiating with lower energy as compared to {\it Year} or vice versa. Since this has no effect on our actual analysis, we have ordered them alphabetically. So, different words which radiate with different energies that appear an equal number of times in this specific Winnie the Pooh story will be classified from lower to higher energy levels alphabetically.

In the column `Energies from data $E(E_i)$', we represent $E(E_i)$, the `amount of energy radiated in the Winnie the Pooh story by the cognitons of a specific word, hence of a specific energy level $E_i$'. The formula for this amount is given by
\begin{eqnarray}
E(E_i) = N(E_i) E_i
\end{eqnarray}
the product of the number $N(E_i)$ of cognitons in the state of the word with energy level $E_i$ multiplied by the amount of energy $E_i$ radiated by such a cogniton in that state.

\small
\begin{longtable}{p{1.5cm}p{1.5cm}p{1.5cm}p{1.5cm}p{1.5cm}p{1.5cm}p{1.5cm}p{1.5cm}}
\label{piglethaffalunmp}
  Words concepts cognitons & Energy levels $E_i$ & Appearance numbers $N(E_i)$ & Bose-Einstein modeling &  Maxwell-Boltzmann modeling & Energies from data $E(E_i)$ & Energies Bose-Einstein & Energies Maxwell-Boltzmann  \\
  \hline
      {\it And}  & 0 & 133 & 129.05 & 28.29 & 0    & 0     & 0  \\
         {\it He} & 1 & 111 & 105.84 & 28.00 & 111 & 105.84 & 28.00  \\
        {\it The} & 2 & 91  & 89.68   & 27.69 & 182 & 179.36 & 55.38  \\
            {\it It} & 3 & 85  & 77.79   & 27.40 & 255 & 233.36 & 82.19  \\
            {\it A} & 4 & 70  & 68.66   & 27.11 & 280 & 274.65 & 108.43  \\
           {\it To} & 5 & 69  & 61.45   & 26.82 & 345 & 307.23 & 234.09  \\
        {\it Said} & 6 & 61  & 55.59   & 26.53 & 366 & 333.55 & 159.20  \\
        {\it Was} & 7 & 59  & 50.75   & 26.25 & 413 & 355.24 & 183.76  \\
      {\it Piglet} & 8 & 47  & 46.68   & 25.97 & 376 & 373.40 & 207.78  \\
              {\it I} & 9 & 46  & 43.20   & 25.70 & 414 & 388.82 & 231.27  \\
        {\it That} & 10 & 41 & 40.21  & 25.42 & 410 & 402.05 & 254.24  \\
       {\it Pooh} & 11 & 40 & 37.59  & 25.15 & 440 & 413.52 & 276.69  \\
         {\it Of}   & 12 & 39 & 35.30  & 24.89 & 468 & 423.55 & 298.64  \\
         {\it Had} & 13 & 28 & 33.26  & 24.62 & 364 & 432.38 & 320.09  \\
      {\it Would} & 14 & 26 & 31.44  & 24.36 & 364 & 440.21 & 341.05  \\
         {\it As}    & 15 & 25 & 29.81  & 24.10 & 375 & 447.19 & 361.53  \\
          {\it In}    & 16 & 25 & 28.34  & 23.86 & 400 & 453.44 & 381.53  \\
        {\it But}    & 17 & 23 & 27.00  & 23.59 & 391 & 459.07 & 401.07  \\
 {\it Haffalump}& 18 & 23 & 25.79  & 23.34 & 414 & 464.15 & 420.15  \\
        {\it His}    & 19 & 23 & 24.67  & 23.09 & 437 & 468.77 & 438.78  \\
      {\it Very}    & 20 & 23 & 23.65  & 22.85 & 460 & 472.96 & 456.97  \\
       {\it You}     & 21 & 23 & 22.70  & 22.61 & 483 & 476.79 & 474.72  \\
      {\it Then}    & 22 & 21 & 21.83  & 22.37 & 462 & 480.30 & 492.05  \\
    {\it Honey}    & 23 & 20 & 21.02  & 22.13 & 460 & 483.51 & 508.95  \\
          {\it So}    & 24 & 20 & 20.27  & 21.89 & 480 & 486.47 & 525.43  \\
          {\it Up}    & 25 & 20 & 19.57  & 21.66 & 500 & 489.19 & 541.51  \\
       {\it They}    & 26 & 19 & 18.91  & 21.43 & 494 & 491.71 & 557.19  \\
             {\it If}    & 27 & 18 & 18.30  & 21.20 & 486 & 494.03 & 572.47  \\
          {\it Jar}    & 28 & 18 & 17.72  & 20.98 & 504 & 496.18 & 587.37  \\
      {\it There}    & 29 & 18 & 17.18  & 20.75 & 522 & 498.18 & 601.89  \\
            {\it At}    & 30 & 17 & 16.67  & 20.53 & 510 & 500.03 & 616.03  \\
           {\it Be}    & 31 & 15 & 16.19  & 20.32 & 465 & 501.75 & 629.80  \\
          {\it Got}    & 32 & 15 & 15.73  & 20.10 & 480 & 503.34 & 643.21  \\
         {\it Just}    & 33 & 15 & 15.30  & 19.89 & 495 & 504.83 & 656.26  \\
        {\it What}    & 34 & 15 & 14.89  & 19.68 & 510 & 506.22 & 668.97  \\
 {\it Christopher} & 35 & 14 & 14.50  & 19.47 & 490 & 507.51 & 681.33  \\
             {\it This} & 36 & 14 & 14.13  & 19.26 & 504 & 508.71 & 693.35  \\
             {\it Trap} & 37 & 14 & 13.78  & 19.06 & 518 & 509.83 & 705.03  \\
           {\it About} & 38 & 13 & 13.44  & 18.85 & 494 & 510.88 & 716.40  \\
                {\it All} & 39 & 13 & 13.12  & 18.65 & 507 & 511.86 & 727.44  \\
         {\it Should} & 40 & 13 & 12.82  & 18.45 & 520 & 512.77 & 738.17  \\
               {\it For} & 41 & 12 & 12.53  & 18.26 & 492 & 513.62 & 748.59  \\
              {\it Like} & 42 & 12 & 12.25  & 18.06 & 504 & 514.41 & 758.70  \\
            {\it Robin} & 43 & 12 & 11.98  & 17.87 & 516 & 515.15 & 768.51  \\
               {\it See} & 44 & 12 & 11.72  & 17.68 & 528 & 515.84 & 778.03  \\
            {\it When} & 45 & 12 & 11.48  & 17.49 & 540 & 516.48 & 778.26  \\
            {\it Down} & 46 & 11 & 11.24  & 17.31 & 506 & 517.08 & 796.20  \\
   {\it Haffalumps} & 47 & 11 & 11.01  & 17.12 & 517 & 517.64 & 804.87  \\
              {\it With} & 48 & 11 & 10.79  & 16.94 & 528 & 518.15 & 813.26  \\
                 {\it Do} & 49 & 10 & 10.58  & 16.76 & 490 & 518.63 & 821.39  \\
                 {\it Go} & 50 & 10 & 10.38  & 16.58 & 500 & 519.08 & 829.25  \\
                 {\it Off} & 51 & 10 & 10.19  & 16.41 & 510 & 519.49 & 836.85  \\
                 {\it On} & 52 & 10 & 10.00  & 16.23 & 520 & 519.87 & 844.19  \\
             {\it Think} & 53 & 10 & 9.82  & 16.06 & 530 & 520.22 & 851.29  \\
         {\it Thought} & 54 & 10 & 9.64  & 15.89 & 540 & 520.54 & 858.13  \\
              {\it More} & 55 & 9   & 9.47    & 15.72 & 495 & 520.83 & 864.74  \\
                  {\it No} & 56 & 9   & 9.31    & 15.56 & 504 & 521.10 & 871.11  \\
                 {\it Out} & 57 & 9   & 9.15    & 15.39 & 513 & 521.35 & 877.25  \\
                   {\it Pit} & 58 & 9   & 8.99    & 15.23 & 522 & 521.57 & 883.15  \\
               {\it Went} & 59 & 9   & 8.84    & 15.07 & 531 & 521.77 & 888.84  \\
               {\it Don't} & 60 & 8   & 8.70    & 14.91 & 480 & 521.95 & 894.30  \\
               {\it Good} & 61 & 8   & 8.56    & 14.75 & 488 & 522.11 & 899.55  \\
               {\it Head} & 62 & 8   & 8.43    & 14.59 & 496 & 522.25 & 904.58  \\
               {\it Know} & 63 & 8   & 8.29    & 14.44 & 504 & 522.37 & 909.41  \\
                   {\it Oh} & 64 & 8   & 8.16    & 14.28 & 512 & 522.48 & 914.03  \\
                {\it Right} & 65 & 8   & 8.04    & 14.13 & 520 & 522.57 & 918.45  \\
                  {\it Well} & 66 & 8   & 7.92    & 13.98 & 528 & 522.64 & 922.67  \\
                  {\it Bed} & 67 & 7   & 7.80    & 13.83 & 469 & 522.70 & 926.70  \\
                {\it Could} & 68 & 7   & 7.69    & 13.68 & 476 & 522.74 & 930.54  \\
                 {\it Deep} & 69 & 7   & 7.58    & 13.54 & 483 & 522.77 & 934.20  \\
                    {\it Did} & 70 & 7   & 7.47    & 13.40 & 490 & 522.78 & 937.67  \\
 {\it \underline {First}} & {\bf \underline {71}} & 7   & 7.36    & 13.25 & 497 & {\bf \underline{522.79}} & 940.96  \\ 
                  {\it Have} & 72 & 7   & 7.26    & 13.11 & 504 & 522.78 & 944.08  \\ 
                   {\it Help} & 73 & 7   & 7.16    & 12.97 & 511 & 522.76 & 947.02  \\ 
               {\it Himself} & 74 & 7   & 7.06    & 12.84 & 518 & 522.72 & 949.79  \\ 
                    {\it How} & 75 & 7   & 6.97    & 12.70 & 525 & 522.68 & 952.40  \\
                {\it Looked} & 76 & 7   & 6.88    & 12.56 & 532 & 522.63 & 954.85  \\
                     {\it Now} & 77 & 7   & 6.79    & 12.43 & 539 & 522.56 & 957.13  \\
                       {\it Put} & 78 & 7   & 6.70    & 12.30 & 546 & 522.49 & 959.27  \\
        \ldots &  \ldots &  \ldots   &  \ldots    & \ldots &  \ldots &  \ldots &  \ldots  \\
                                   &     &      &      &     &      &      &       \\
                                   &     &      &      &     &      &      &       \\
       \ldots &  \ldots &  \ldots   &  \ldots    & \ldots &  \ldots &  \ldots &  \ldots  \\
                 {\it Wishing} & 538 & 1   & 0.67    & 0.09 & 538 & 359.92 & 48.65  \\
                     {\it Word} & 539 & 1   & 0.67    & 0.09 & 539 & 359.58 & 48.22  \\
                   {\it Worse} & 540 & 1   & 0.67    & 0.09 & 540 & 359.24 & 47.80  \\
                      {\it Year} & 541 & 1   & 0.66    & 0.09 & 541 & 358.90 & 47.38  \\
                   {\it You've} & 542 & 1   & 0.66    & 0.09 & 542 & 358.55 & 46.96  \\
                   \hline
                                     &      &2655   &2655.00    &2654.96 & 242891 & 242891.01 & 242889.76  \\
\caption{{\footnotesize An energy scale representation of the words of the Winnie the Pooh story `In Which Piglet Meets a Haffalump' by A. A. Milne as published in \citet{milne1926}. The words are in the column `Words concepts cognitons' and the energy levels are in the column `Energy levels $E_i$', and are attributed according to the `numbers of appearances' in the column `Appearance numbers $N(E_i)$', such that  lower energy levels correspond to a higher order of appearances, and the value of the energy levels is determined according to (\ref{Ei}). The `amounts of energies radiated by the words of energy level $E_i$' are in the column `Energies from data $E(E_i)$'. In the columns `Bose--Einstein modeling', `Maxwell--Boltzmann modeling', `Energies Bose--Einstein', and `Energies Maxwell--Boltzmann' are respectively the predicted values of the Bose--Einstein and the Maxwell--Boltzmann model of the `numbers of appearances', and of the `radiated energies'. In the graphs of Figure \ref{piglethaffalunmpenergygraph}, we can see that a maximum is reached for the energy level $E_{71}$, corresponding to the word {\it First}, which appears seven times in the Winnie the Pooh story. If we use the analogy with light, we can say that the radiation spectrum of the story `In Which Piglet Meets a Haffalump' has a maximum at {\it First}, which would hence be, again in analogy with light, the dominant color of the story. We have indicated this radiation peak by underlining the word {\it First}, its energy level $E_{71}$, and the amount of energy 522.79 the story radiates, following the Bose-Einstein model. The omitted words are those between the word {\it Put} at number 78, and the word {\it Wishing} at 538, and this set of omitted words are indicated by a blank space in the Table. }}
\end{longtable}
\normalsize
\noindent
In the last row of Table \ref{piglethaffalunmp}, we give the {\it Totalities}, namely in the column `Appearance numbers $N(E_i)$' of this last row the total number of words 
\begin{eqnarray}
\sum_{i=0}^n N(E_i) = N = 2655
\end{eqnarray}
{and in the colu}mn `Energies from data $E(E_i)$' of the last row we give the total amount of energy 
\begin{eqnarray}
\sum_{i=0}^n E(E_i) = \sum_{i=0}^n N(E_i) E_i = E = 242{,}891
\end{eqnarray}
radiated by the Winnie the Pooh story `In Which Piglet Meets a Haffalump'. Hence, the columns `Words concepts cognitons', `Energy levels $E_i$', `Appearance numbers $N(E_i)$', and `Energies from data $E(E_i)$' contain all the experimental data of the Winnie the Pooh story `In Which Piglet Meets a Haffalump'.

In columns `Bose--Einstein modeling' and `Maxwell--Boltzmann modeling' of Table \ref{piglethaffalunmp}, we give the values of the populations of the different energy states for, respectively, a Bose--Einstein and a Maxwell--Boltzmann model of the data of the considered story. Let us explain what these two models are.
 
The Bose--Einstein distribution function is given by
\begin{eqnarray} \label{boseeinsteindistribution}
N(E_i) = {1 \over {Ae^{{E_i \over B}}-1}}
\end{eqnarray}
where $N(E_i)$ is the number of bosons obeying the Bose--Einstein statistics in energy level $E_i$ and $A$ and $B$ are two constants  that are determined by expressing that the total number of bosons equals the total number of words, and that the total energy radiated equals the total energy of the Winnie the Pooh story `In Which Piglet Meets a Haffalump', hence by the two conditions
\begin{eqnarray} \label{BEconstraint}
\sum_{i=0}^n {1 \over {Ae^{{E_i \over B}}-1}} = N = 2655 \quad {\rm and} \quad \sum_{i=0}^n {E_i \over {Ae^{{E_i \over B}}-1}} = E = 242{,}891
\end{eqnarray}

We remark that the Bose--Einstein distribution function is derived in quantum statistical mechanics for a gas of bosonic quantum particles where the notions of `identity and indistinguishability' play the specific role they are attributed in quantum theory \citep{huang1987,yokoiabe2018}.

Since we want to show the validity of the Bose--Einstein statistics for words in human language, we compared our Bose--Einstein distribution model with a Maxwell--Boltzmann distribution model, hence we introduce also the Maxwell--Boltzmann distribution explicitly. It is the distribution described by the following function 
\begin{eqnarray} \label{maxwellboltzmanndistribution}
N(E_i) = {1 \over Ce^{{E_i \over D}}}
\end{eqnarray} 
where $N(E_i)$ is the number of classical identical particles obeying the Maxwell--Boltzmann statistics in energy level $E_i$, and $C$ and $D$ are two constants that will be determined, like in the case of the Bose--Einstein statistics, by the two conditions
\begin{eqnarray} \label{MBconstraint}
\sum_{i=0}^n {1 \over Ce^{{E_i \over D}}} = N = 2655 \quad {\rm and} \quad \sum_{i=0}^n {E_i \over Ce^{{E_i \over D}}} = E = 242{,}891
\end{eqnarray}

The Maxwell--Boltzmann distribution function is derived for `classical identical and distinguishable' particles, and can also be shown in quantum statistical mechanics to be a good approximation if the quantum particles are such that their `de Broglie waves' do not overlap  \citep{huang1987}. In the last two columns of Table \ref{piglethaffalunmp}, `Energies Bose--Einstein' and `Energies Maxwell--Boltzmann', we show the `energies' related to the Bose--Einstein modeling and to the Maxwell--Boltzmann modeling, respectively.

We have now introduced all that is necessary to announce the principal result of our investigation in \citet{aertsbeltran2020}. 
\begin{quotation}
\noindent
When we determine the two constants $A$ and $B$, respectively $C$ and $D$, in the Bose--Einstein distribution function (\ref{boseeinsteindistribution}) and Maxwell--Boltzmann distribution function (\ref{maxwellboltzmanndistribution}), by putting the total number of particles of the model equal to the total number of words of the considered text and by putting the total energy of the model to the total energy of the considered text, (\ref{BEconstraint}) and (\ref{MBconstraint}), we find a remarkably good fit of the Bose--Einstein modeling function with the data of the text, and a big deviation of the Maxwell--Boltzmann modeling function with respect to the data of the text. 
\end{quotation}

\begin{figure}
    \centering
    \subfloat[Numbers of appearances distribution graphs]{{\includegraphics[width=8cm]{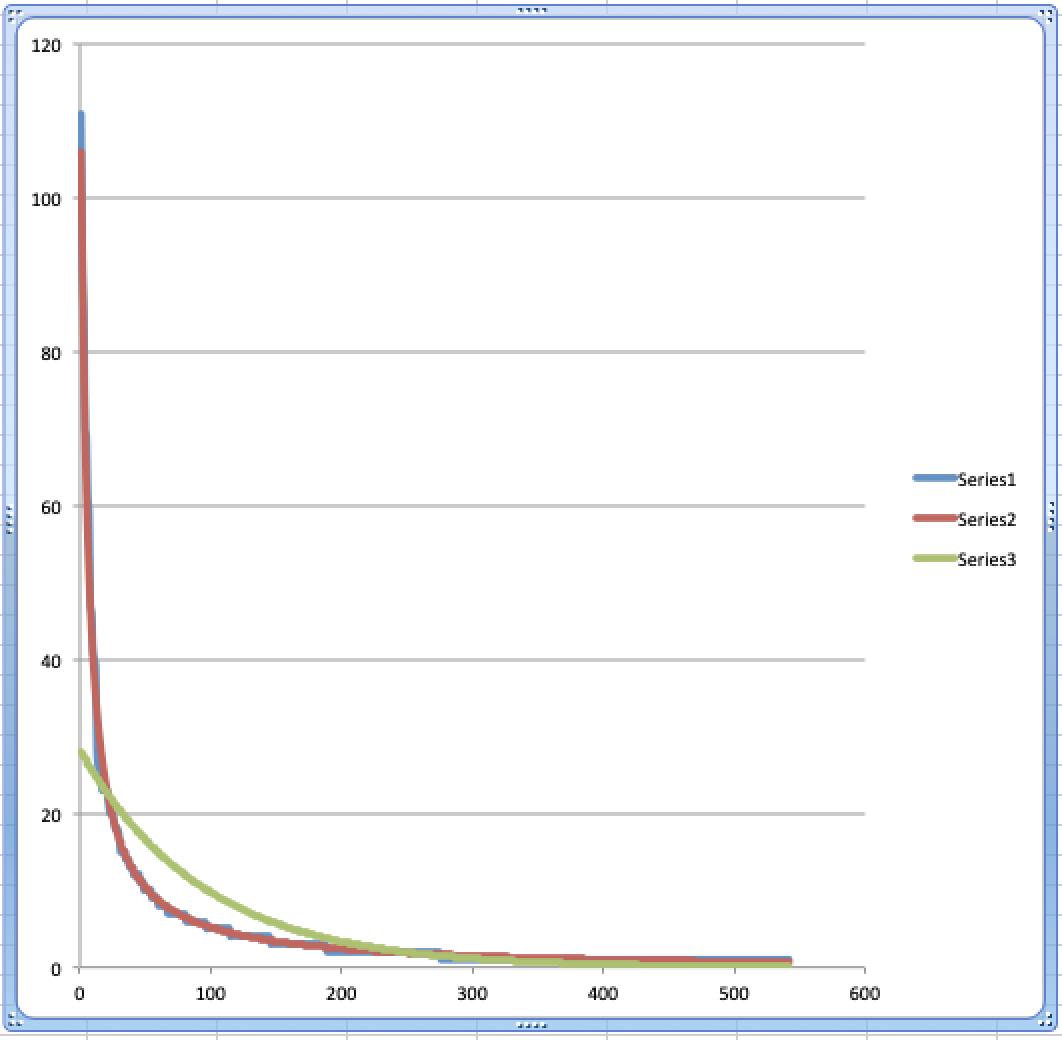} }}%
    \qquad
    \subfloat[$\log/\log$ graphs of numbers of appearances distributions]{{\includegraphics[width=8cm]{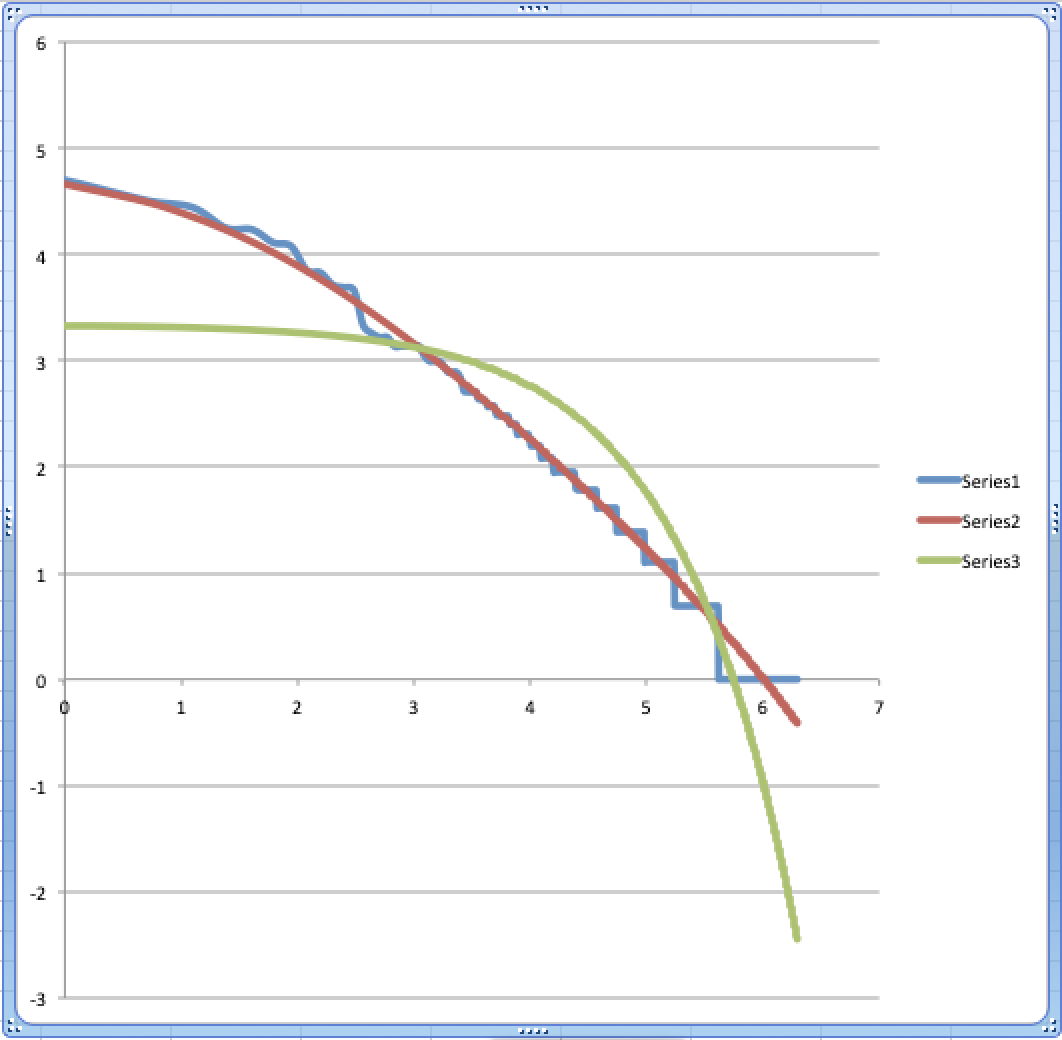} }}%
    \caption{In (\textbf{a}) we represent the `number of appearances' of words in the Winnie the Pooh story `In Which Piglet Meets a Haffalump' \citep{milne1926}, ranked from lowest energy level, corresponding to the most often appearing word, to highest energy level, corresponding to the least often appearing word as listed in Table \ref{piglethaffalunmp}. The blue graph (Series 1) represents the data, i.e. the collected numbers of appearances from the story (column `Appearance numbers $N(E_i)$' of Table \ref{piglethaffalunmp}), the red graph (Series 2) is a Bose--Einstein distribution model for these numbers of appearances (column `Bose--Einstein modeling' of Table \ref{piglethaffalunmp}), and the green graph (Series 3)  is a Maxwell--Boltzmann distribution model (column `Maxwell--Boltzmann modeling' of Table \ref{piglethaffalunmp}). In (\textbf{b}) we represent the $\log / \log$ graphs of the `numbers of appearances' and their Bose--Einstein and Maxwell--Boltzmann models. The red and blue graphs coincide almost completely in both (\textbf{a},\textbf{b}) whereas the green graph does not coincide at all with the blue graph of the data. This shows that the Bose--Einstein distribution is a good model for the numbers of appearances, while the Maxwell--Boltzmann distribution is not.}%
    \label{piglethaffalunmpgraphpiglethaffalunmploggraph}%
\end{figure}

The result is expressed in the graphs of Figure \ref{piglethaffalunmpgraphpiglethaffalunmploggraph}a, where the blue graph represents the data, hence the numbers in column `Energies from data $E(E_i)$' of Table \ref{piglethaffalunmp}, the red graph represents the quantities obtained by the Bose--Einstein model, hence the quantities in column `Bose--Einstein modeling' of Table \ref{piglethaffalunmp}, and the green graph represents the quantities obtained by the Maxwell--Boltzmann model, hence the quantities of column `Energies Maxwell--Boltzmann' of Table \ref{piglethaffalunmp}. We can easily see in Figure \ref{piglethaffalunmpgraphpiglethaffalunmploggraph}a how the blue and red graphs almost coincide, while the green graph deviates abundantly from the two other graphs which show how Bose--Einstein statistics is a very good model for the data we collected from the Winnie the Pooh story, while Maxwell--Boltzmann statistics completely fails to model these data.

\begin{figure}
\centering
\includegraphics[width=8cm]{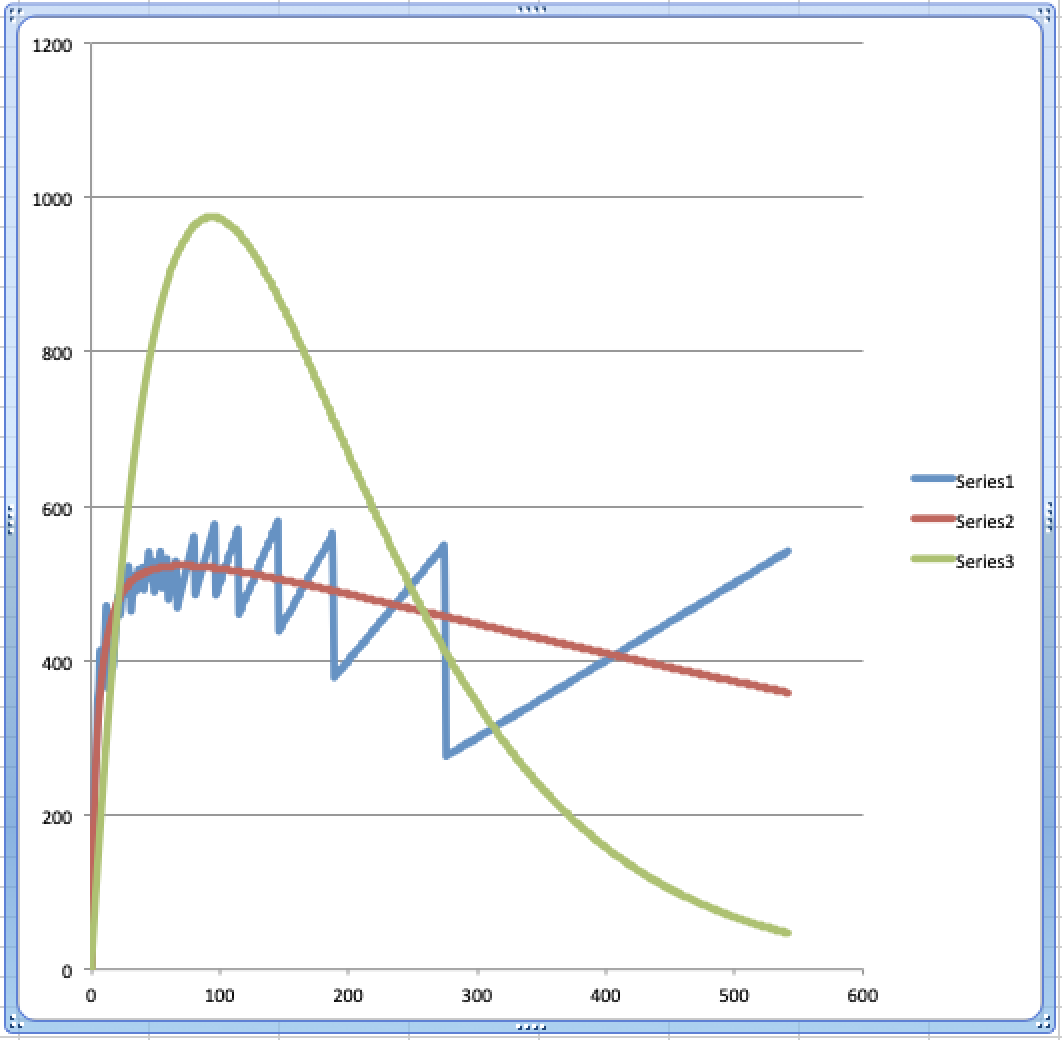}
\caption{A representation of the `energy distribution' of the Winnie the Pooh story `In Which Piglet Meets a Haffalump' \citep{milne1926} as listed in Table \ref{piglethaffalunmp}. The blue graph (Series 1) represents the energy radiated by the story per energy level (column `Energies from data $E(E_i)$' of Table \ref{piglethaffalunmp}), the red graph (Series~2) represents the energy radiated by the Bose--Einstein model of the story per energy level (column `Energies Bose--Einstein' of Table \ref{piglethaffalunmp}), and the green graph (Series 3) represents the energy radiated by the Maxwell--Boltzmann model of the story per energy level (column `Energies Maxwell--Boltzmann' of Table \ref{piglethaffalunmp}).}
\label{piglethaffalunmpenergygraph}
\end{figure}

We also considered the energies, and following the constraints (\ref{BEconstraint}) and (\ref{MBconstraint}), the total energy of the Bose--Einstein model and the total energy of the Maxwell--Boltzmann model are both equal to the total energy of the data of the Winnie the Pooh story. The result of both constraints  (\ref{BEconstraint}) {and} (\ref{MBconstraint}) on the energy functions that express the amount of energy per energy level can be seen in Figure \ref{piglethaffalunmpenergygraph}. We see again that the red graph, which represents the Bose--Einstein radiation spectrum, is a much better model for the blue graph, which represents the experimental radiation spectrum, as compared to the green graph, which represents the Maxwell--Boltzmann radiation spectrum. 

Both solutions, the Bose--Einstein shown in the red graph and the Maxwell--Boltzmann shown in the green graph, have been found by making use of a computer program calculating the values of $A$, $B$, $C$, and $D$ such that (\ref{BEconstraint}) and (\ref{MBconstraint}) are satisfied, which gives the approximate values
\begin{eqnarray} \label{gastemperature}
A  \approx 1.0078 \quad B  \approx 593.51 \quad C  \approx 0.0353 \quad D \approx 93.63
\end{eqnarray}

In the graphs of Figure \ref{piglethaffalunmpenergygraph}, we can see that a maximum is reached for the energy level $E_{71}$, corresponding to the word {\it First}, which appears seven times in the Winnie the Pooh story. As it will be analyzed very explicitly in Section \ref{identifyindistinguishability}, we make a comparison between `a text and its words' and `electromagnetic radiation and its photons', where the words and the photons are the quanta of the text and the electromagnetic radiation, respectively. 
The word {\it First} is then the word that emits the most radiation, namely the amount of energy equal to $522.79$. The words that come before it, with a lower energy level, emit less radiation, and the words that come after it, with a higher energy level, also emit less radiation, because, although they individually belong to a higher energy level, they become smaller in number, so that the effect of the higher energy level is cancelled out by this smaller number when calculating the total emitted energy per energy level.
That is why we have underlined this word {\it First} in the Table \ref{piglethaffalunmp}, and we have also underlined the energy level in which it is found, namely $E_{71}$, as well as the amount of radiation it emits. Using again the analogy with electromagnetic radiation, i.e. with light, the word {\it First}, and the words close to it in terms of energy level, determine the `color' of the story of Winnie the Pooh `In Which Piglet Meets a Haffalump'. From the course of the graphs in Figure \ref{piglethaffalunmpenergygraph}, we can note this maximum, and consider the similarity of the graphs in Figure \ref{piglethaffalunmpenergygraph} to the well-known curve of the radiation energy of light, as described by Planck's radiation law \citep{planck1900} (see (\ref{planckradiation})).

 Due to their shape, the graphs in Figure \ref{piglethaffalunmpgraphpiglethaffalunmploggraph}a are not easily comparable, and although quite obviously the blue and red graphs are almost overlapping, while the blue and green graphs are very different, which shows that the data are well modeled by Bose--Einstein statistics and not well modeled by Maxwell--Boltzmann statistics, it is interesting to consider 
a transformation where we apply the $\log$ function to both the $x$-values, i.e. the domain values, and the $y$-values, i.e. the image values, of the functions underlying the graphs. This is a well-known technique to render functions giving rise to this type of graphs more easily comparable. 

In Figure \ref{piglethaffalunmpgraphpiglethaffalunmploggraph}b, the graphs can be seen where we have taken the $\log$ of the $x$-coordinates and also the $\log$ of the $y$-coordinates of the graph representing the data, which is again the blue graph in Figure \ref{piglethaffalunmpgraphpiglethaffalunmploggraph}b, of the graph representing the Bose--Einstein distribution model of these data, which is the red graph in Figure \ref{piglethaffalunmpgraphpiglethaffalunmploggraph}b, and of the graph representing the Maxwell--Boltzmann distribution model of the data, which is the green graph in Figure \ref{piglethaffalunmpgraphpiglethaffalunmploggraph}b. 

Readers acquainted with the well-known Zipf's law as it appears in human language \citep{zipf1935,zipf1949} will recognize Zipf's graph in the blue graph of Figure \ref{piglethaffalunmpgraphpiglethaffalunmploggraph}b. It is indeed the $\log/\log$ graph of `ranking' versus 'numbers of appearances' of the text of the Winnie the Pooh story `In Which Piglet Meets a Haffalump' which is the `definition' of Zipf's graph. As to be expected, we see Zipf's law being satisfied, the blue graph is well approximated by a straight line with a negative gradient close to $-$1. We see that the Bose--Einstein graph still models very well this Zipf's graph, and what is more, it also models the (small) deviation from Zipf's graph of the straight line. Zipf's law and the corresponding straight line when a $\log/\log$ graph is drawn is an empirical law. Intrigued by the modeling of the Bose--Einstein statistics by the Zipf graph, we have analyzed this correspondence in more detail in \citet{aertsbeltran2020}, but definitely plan in future work to investigate further this correspondence between our Bose--Einstein energetic model and Zipf's experimental law in human language.

\section{Photons, atoms and words \label{identifyindistinguishability}}
Let us consider Max Planck's black body radiation law \citep{planck1900},
\begin{eqnarray} \label{planckradiation}
B(\nu,T) = {2 h \nu^3 \over c^2} {1 \over {e^{h\nu \over kT} - 1}}
\end{eqnarray} 
and identify the different elements. $B(\nu,T)$ is the amount of energy, per unit surface area, per unit time, per unit solid angle, per unit frequency at frequency $\nu$. $T$ is the absolute temperature, $h$ is Planck's constant, $c$ is the speed of light, and $k$ is Boltzmann's constant. Let us also consider Wien's approximation to Planck's law, 
\begin{eqnarray} \label{wienradiation}
B(\nu,T)_{Wien} = {2 h \nu^3 \over c^2} {1 \over {e^{h\nu \over kT}}}
\end{eqnarray}

We can see that when substituting the Bose--Einstein statistical part of (\ref{boseeinsteindistribution}) in (\ref{planckradiation}) with the Maxwell--Boltzmann statistical part of (\ref{maxwellboltzmanndistribution}), we get (\ref{wienradiation}), and indeed, Wien's approximation is a Maxwell--Boltzmann approximation of Planck's law. Historically, Wien's law was formulated before Planck's law \citep{wien1897}, and it gave a good approximation for high frequency radiation (if $h\nu >> kT$ we can indeed approximate $1 / (e^{h\nu \over kT} - 1)$ well by ${1 / {e^{h\nu \over kT}}}$), it was not Wien who introduced the two constants, $h$ and $k$, that was Planck, Wien proposed an arbitrary constant in the law as he formulated it. In the years that Planck dedicated to the study of black body radiation, new experiments had shown a serious deviation from Wien's law for low-frequency radiation. This motivated Planck to propose a new law for black body radiation, introducing the notion of `quantization' into physics and initiating the period called the `Old Quantum Theory'. 

Because it is illuminating for the hypothesis we put forward in this article, namely that words are the `quanta of human language', we will now outline some historical aspects of the Old Quantum Theory. More specifically, we want to make clear that the situation which was to ultimately lead to the Bose--Einstein quantum statistics was already identified and abundantly discussed between 1905, the date of Einstein's publication on the photo-electric effect \citep{einstein1905}, and 1924, the date of Einstein's articles containing the prediction of the Bose--Einstein condensate \citep{einstein1924,einstein1925a,einstein1925b}. Additionally, we want to make clear that the focus of this discussion centered primarily on the specific derivation developed by Max Planck of the radiative law, grounded in Boltzmann's theory of the statistical aspects of thermodynamics. This period has since been thoroughly studied by philosophers of science and historians of science, on the basis of the original German-language articles and communications of scientists, and our brief summary of it is based on the following articles \citep{howard1990,monaldi2009,perezsauer2010,gorroochurn2018}.
 
The young Einstein was the first to make explicit use of Planck's hypothesis when in 1905 he proposed for the photoelectric effect a description that considered the energy packets as light quanta much more explicitly than was Planck's intention \citep{einstein1905}. At the time, Paul Ehrenfest, a student of Ludwig Boltzmann's and a close friend of Einstein's, was, next to Einstein and Planck, most concerned with the new hypothesis about light quanta. Both Einstein and Ehrenfest increasingly wanted to consider the light quanta as entities in their own right, with, for example, an independent existence in space, something Planck was not at all in favor of. There was also a growing understanding that when considering the light quanta as entities in themselves, like atoms and molecules of a gas, Planck's derivation of the radiation formula exposed a fundamental problem. The probabilities for the relevant microstates, which were necessary for Boltzmann's theory in the calculation of entropy, did not appear to be statistically independent for the different light quanta. More to the point, if the probabilities for the microstates for different light quanta were assumed to be independent, the radiation law again became that of Wien, Ehrenfest deduced \citep{ehrenfest1911,monaldi2009}. In the period following this proof by Ehrenfest, Einstein nevertheless remained faithful to his view of the light quanta as entities existing in space by themselves, translating statistical dependence into the existence of correlations, and this notion of light quanta was gaining ground among other scientists. For some, the idea that light particles formed in conglomerates of light molecules began to make sense. However, there were also numerous physicists who followed the view of Planck, and explained the peculiar statistics of the energy elements required by Planck's law of radiation as being rather useful `fictions' of the `resonator model', and awaiting further discoveries related to it \citep{howard1990,perezsauer2010}. 
 
In this respect, let us not forget that even in his original analysis Planck was inspired by a similarity between `the radiation of a black body' and `the behavior of an ideal gas', in that he used Boltzmann's theory, developed primarily for gases, for his radiation law. Einstein and Erhrenfest were captivated by the same analogy, and the differences of opinion were due to the peculiar statistical dependence that Planck's formula required, whereas, for Wien's formula, the Maxwell--Boltzmann approximation to Planck, that peculiar dependence did not arise \citep{monaldi2009,perezsauer2010}. Things took a new turn when Einstein received a letter from a young Indian scientist, Satyendra Nath Bose, who claimed to have found a new derivation of Planck's radiation law. Bose had tried to publish his paper but had failed, and he sent it to Einstein, suspecting that the latter would be interested, Bose was indeed aware of the discussions about Planck's radiation law that were going on. Einstein was indeed intrigued, he translated Bose's article into German and promptly sent it to Zeitschrift f\"ur Physik, where it was published \citep{bose1924}. However, more, immediately he set to work introducing Bose's approach for a model of an ideal gas, about which he published an article the same year, and two additional articles in January of the following year \citep{einstein1924,einstein1925a,einstein1925b}.
  
Ehrenfest was not at all enthusiastic about Bose's method and made this clear to a fellow physicist, writing the following in a letter dated October 9, 1924, to Abram Joff\'e: ``Precisely now Einstein is with us. We coincide fully with him that Bose's disgusting work by no means can be understood in the sense
that Planck's radiation law agrees with light atoms moving independently (if they move independently one of each other, the entropy of radiation would depend on the volume not as in Planck, but as in W. Wien, i.e. in the following way: $\kappa \log V^{E \over h}$).'' (in \citet{moskovchenkofrenkel1990}, pp. 171--172). Einstein was silent on the problem of the statistical dependence of the light quanta leading to Planck's law of radiation using Bose's method in his first quantum gas article \citep{einstein1924}. The second article \citep{einstein1925a} he wrote during a stay in Leiden with Ehrenfest, and in it he openly states, ``Ehrenfest and other colleagues have objected, regarding Bose's theory of radiation and my analogous theory of the ideal gas, that in these theories the quanta or molecules are not treated as statistically independent entities, without this circumstance being specifically pointed out in our articles. This is entirely correct." (in \citet{einstein1925a} page 5).  

It is important to note that, parallel to these events, Louis de Broglie had come forward with the hypothesis of `a wave character also for material particles' \citep{debroglie1924}, and, although Einstein only learned of de Broglie's writings after his contacts with Bose, he had read de Broglie's thesis with great interest before writing his second article on the quantum gas~\citep{einstein1925a}. In this article, it becomes clear that he sees the statistical dependence of gas atoms as caused by a `mutual influence', the nature of which is as yet totally unknown. However, he also mentions that the wave-particle duality, brought out in de Broglie's work, might play a role in unraveling it \citep{einstein1925a,monaldi2009}. Not much later, in a letter to Erwin Schr\"odinger, who had begun to take an interest in the quantum gas at this time \citep{schrodinger1926a}, Einstein wrote the following passage: ``In Bose's statistics, which I have used, the quanta or molecules are not treated as independent from one another. [ \ldots ] According to this procedure, the molecules do not seem to be localized independently from each other, but they have a preference to be in the same cell with other molecules. [\ldots ] According to Bose, the molecules crowd together relatively more often than under the
hypothesis of statistical independence.'' (in~\citet{monaldi2009} page~392, see also Figure 1 in \citet{monaldi2009}, where this passage is shown in Einstein's handwriting). So here Einstein explained in a crystal clear way the nature of this statistical dependence and opened the door for the introduction of a new statistics, the Bose--Einstein statistics.
Several letters between Einstein and Schr\"odinger in the late year of 1925 about Einstein's quantum gas, and in parallel Schr\"odinger taking note of de Broglie's work on the quantum theory of gases \citep{debroglie1924}, were instrumental in Schr\"odinger's formulation of wave mechanics \citep{schrodinger1926b,howard1990}. When Einstein realized what it meant that Schr\"odinger's matter waves were defined on the configuration spaces of the quanta, which made these waves fundamentally different from de Broglie's waves, it marked the turning point for him, where he would step away from the new more and more abstract mathematically defined quantum formalism \citep{howard1990}.  

Let us now show how  this statistical dependence of the different quanta, which has been the subject of debate among the scientists of the Old Quantum Theory for so many years, occurs for words, the quanta of human language. First, we illustrate what this statistical dependence means for a very simple situation, the situation of two particles and two states. If we consider this situation of two particles and two states in the manner of Maxwell--Boltzmann, four distinguishable situations occur, (1) the two particles are both in the first state, (2) the two particles are both in the second state, (3) the first particle is in the first state and the second particle is in the second state, and (4) the first particle is in the second state and the second particle is in the first state. These four cases are realized, each by itself, with probability 1/4, as a consequence of the statistical independence of the two particles of the states in which they may be. A Bose--Einstein way of looking at this situation considers only three cases, (1) the two particles are both in the first state, (2)~the two particles are both in the second state, (3) one particle is in the first state and the other particle is in the second state. Now probability 1/3 is assigned to each of these three cases in itself. Some reflection on this simple situation makes it clear why assigning 1/3 probability to each of these cases separately `cannot' be compatible with the independent behavior of the individual particles. One might make the mistake of thinking that the indistinguishability can be interpreted as an epistemic indistinguishability, i.e. we cannot distinguish the two cases (3) and (4) in the Maxwell--Boltzmann view experimentally, even though they are ontologically different, but this is not correct. With such epistemic-only indistinguishability, the probabilities would not become 1/3 for each of the separate cases in the Bose--Einstein view, they would become 1/4, 1/4, and 1/2. So there is something else going on than just epistemic indistinguishability. If we compare 1/3, 1/3, and 1/3, the Bose--Einstein probabilities, with 1/4, 1/4, and 1/2, the epistemic indistinguishability Maxwell--Boltzmann probabilities, we see that Bose--Einstein assigns `a higher probability' to cases (1) and (2), i.e. both particles are in the first state, and, both particles are in the second state. Hence the expression that with Bose--Einstein, `particles clump together in the same state, more than one would expect'. Even the expression `Bose--Einstein condensate' arose from this view of `clumping together in the same state', in this case the state of lowest energy. On the other hand, case (3), which in the epistemic indistinguishability Maxwell--Boltzmann view should be assigned probability 1/2, in Bose--Einstein is assigned a smaller probability, namely 1/3. Thus, Bose particles, in addition to their tendency to clump together in the same states, also tend to flee from being in different states.

We now wish to show that this peculiar statistical dependence, which, as we now know from our historical description, was already unintentionally introduced when Max Planck wrote down his new radiation law for a black body, and which was fiercely debated by scientists of the Old Quantum Theory, can be understood and explained in respect of how it occurs in human language. In human language, both `the clumping of words together in the same state', and `the fleeing of words from different states', can be understood and explained because human language is the carrier of `meaning'. Before proceeding to the first new insight we wish to bring forward in this article, let us explain some additional aspects of our quantum prototype theory for concepts and for words \citep{aertsgabora2005a,aertsgabora2005b}. Our approach answers in a specific but not ambiguous way many questions one might ask, for example, the question, `carry words or concepts present in different places in a text the same meaning?'. Let us give an example. The concept {\it Animal} corresponding to the word `animal' in the `Example Text', where it appears in the sentence `From the small grove of trees growing a little further along the riverbank, another animal appears, surveying the panorama', is in a different state than the concept {\it Animal} where it corresponds to the word `animal' where it appears in the sentence `The animal saunters slowly down toward the great river'. In the two cases, the word `animal' is surrounded by a different cloud of other words, and this affects a local state of the concept {\it Animal} and also its local meaning. If the meaning of the whole text is considered, this will again change the state of the concept {\it Animal}, now corresponding to the different places where the word `animal' appears in the text. It means that our quantum version of Rosch's prototype theory allows a relation between `meaning' and `state of a concept' which is similar to the relation between `coherence' and `state of a quantum entity'. Not only words but also symbols can indicate concepts, which is for example often the case in mathematics, hence the notion of `concept' is broader than the notion of `word'. One might get the impression, considering this fluidity of the way words and concepts are approached in our quantum prototype theory, that this is too complex when compared to the role that photons play as particles of light, and atoms or molecules as particles of a boson gas. Indeed, we are still used to thinking, for example, that a photon always possesses a well-defined color, i.e. a well-defined energy or frequency. However, photons in principle exist in all possible `superpositions' and `entanglements' of states with well-defined colors and are now even literally being created in such superposition and entanglement states as a function of the development of quantum computing \citep{kuesetal2017}. In this sense, it is no coincidence that, in \citet{aertsbeltran2020}, we compared the human language to a boson gas close to absolute zero, in its transition to a Bose--Einstein condensate. At that temperature, the superpositions and entanglements in Fock space are so overlapping that there are hardly any individual bosons.  

Let us now proceed to show that we can explain the behavior of words of human language as atoms or molecules of a Bose gas and as photons of light. More precisely, we shall thus explain why words behave under that peculiar statistical dependence noted by scientists of the Old Quantum Theory, as it occurs in Bose--Einstein statistics.  Additionally, as we already announced, our explanation will be connected to human language being the carrier of meaning when used to tell or write down stories. For this, we consider again the example of the Winnie the Pooh story entitled `In Which Piglet Meets a Haffalump' analyzed in Section \ref{wordsasquanta}.  Let us look in Table \ref{piglethaffalunmp}, and as the first word we choose `Piglet', and we see that it appears 47 times in the story. As the second word, we consider `himself', and again we look in Table \ref{piglethaffalunmp}, and there we see that it occurs seven times in the story. Now suppose that it would be possible to ask Alan Alexander Milne to write a few extra paragraphs for the story. From the nature of what the story carries as meaning, it will then be substantially more likely for the word `Piglet' to reappear in the added paragraphs, than it is for the word `himself'. What does this mean? It means that `meaning' has as a consequence that `equal words clump together', and thus give rise to a Bose--Einstein-like statistics, where the quanta, i.e. words, in this case, do not independently choose different states. The probability of a state that is `closer to the meaning of the whole story' being chosen is higher than that of a state that is further away from the meaning of the story. Let us note that we interpret the notion of `meaning' very broadly in the above assertion. Indeed, many of the words that appear frequently in Table \ref{piglethaffalunmp} do not carry much meaning in the narrow sense of the word. They are there primarily to give `structure' to the text, to form, as it were, `the canvas on which the words that explicitly carry meaning play their part'. However, this canvas and its structure are essential so that the explicit form of meaning can also see life. To make this clear with a metaphor, if the text is a play, then the costume designer and even the ticket seller at the entrance also contribute to the play, this is the broad way we interpret `meaning' carried in a text.

Let us give one more example to clarify the above from a related, but slightly different angle. Suppose we consider the concepts {\it Cat} and {\it Dog} and the configurations {\it Two Cats}, {\it Two Dogs} and {\it A Cat And A Dog}. Let us remark that this is exactly the situation we have studied already in great detail showing Bose--Einstein statistics to be a better representation as compared to Maxwell--Boltzmann statistics \citep{aerts2009a,aertssozzoveloz2015b,beltran2021}. Let us first note that, as we explained above, both {\it Cat} and {\it Dog} are states of the concept {\it Animal}, and so here we consider the situation of two particles {\it Animal}, in two states {\it Cat} and {\it Dog}, which is the situation we analyzed theoretically above.
We first consider a pure Maxwell--Boltzmann situation with respect to these three cases. Suppose we visit a farm with a lot of `animals', all of them `cats' or `dogs' living at the farm, more or less equal in number, and we receive as a present two of them randomly chosen for us by the farmer. Then the chance that the gift will be a cat and a dog will be twice that of it being either two cats or two dogs. Hence, the probability distribution of the three cases will be 1/4 for two cats, 1/4 for two dogs, and 1/2 for a cat and a dog. Indeed, to us, as receivers of the present, there is no difference between `a cat and a dog' or `a dog and a cat'. Epistemologically, within the reality they will live in with us, the order has no importance. Even so, the situation that we have called the epistemic Maxwell--Boltzmann situation of indistinguishability gives rise to the probabilities 1/4, 1/4, and 1/2, and not to the Bose--Einstein probabilities 1/3, 1/3, and 1/3. What, however, if we asked a child that has been promised they can have two pets and choose for themselves whether either pet is a cat or a dog. The microstates that come into play in this case exist in the conceptual realm of the child's conceptual world, and there is no reason that within this conceptual world there will be a double amount of microstates for the choice of a cat and a dog as compared to the choices for two cats or two dogs. Specifically this type of situation was investigated by us in many different and more complex configurations, with the result of Bose--Einstein proving a better statistics than Maxwell--Boltzmann to model the situation \citep{aerts2009a,aertssozzoveloz2015b,beltran2021}. If there are however two children each choosing one pet and doing so independently of each other, Maxwell--Boltzmann statistics will be the governing statistics, because the choices of the two children will both carry probabilities 1/2 for cat and 1/2 for dog, which leads to 1/4 for two cats, 1/4 for two dogs, and 1/2 for a cat and a dog. This example shows that the presence of the `clumping of equals' not only occurs in human language, but is also a property of human cognition. There, too, this clumping together is carried by the presence of meaning. Two cats, two dogs, a cat and a dog, are rather treated on an equal footing by the human mind, thus with probabilities 1/3, 1/3, and 1/3, instead of on Maxwell--Boltzmann's, where twice as much probability should be attributed to `a cat and a dog' compared to the probability attributed to `two cats' and `two~dogs'.

Our brief historical presentation of the search for a better understanding of Planck's law of radiation shows how it was characterized by taking seriously the similarity between `the electromagnetic field and its photons' and `an ideal boson gas and its atoms or molecules', and how Albert Einstein, in particular, was guided by it again and again, as new obstacles appeared. We wish to note that in our work in \citet{aertsbeltran2020}, and in the present article, we are guided in a similarly inspired way by a similarity of human language to both the electromagnetic field and its photons, and an ideal boson gas and its atoms or molecules. It is also in this sense, as we already noted in Section \ref{introduction}, that we situate our research within the research domain of quantum cognition and quantum information science. Although it is not our focus in this article, we wish to mention that the new finding, namely how `meaning' gives rise to the Bose--Einstein type of 'clumping of the same words', hence the lack of statistical independence of such words, is a support in a specific way for our `conceptuality interpretation' of quantum mechanics. In the `conceptuality interpretation', quantum entities are assumed to be `concepts' and not `objects' \citep{aerts2009b,aerts2010a,aerts2010b,aerts2013,aertsetal2018d,aertsetal2019b}. In the archetypal situation where a measurement apparatus interacts with a quantum entity, the measurement apparatus plays with respect to the quantum entity the equivalent role that the human mind plays with respect to the human concepts. We should note at the outset that when we claim that quantum entities are `concepts' and not `objects', we do not mean that they are `human concepts'. On the contrary, just like human concepts are the building blocks of human language, so these quantum concepts are the building blocks of the quantum fields, which thus behave like quantum languages within the conceptuality interpretation. For example, the photons are then the concepts, i.e. words, of the electromagnetic field, which behaves like a language. Thus, it is a question of equivalence in behavior and not equality, in the sense that, for example, sound waves and electromagnetic waves are both `waves', but are indeed not the same entities. 

The role of `meaning' for human language is played by `quantum coherence' for the quantum language. One might ask the question, if the quantum concepts, i.e. the quantum Bose particles in the conceptuality interpretation, are not human concepts, then what are they? Since, at least for now, we are talking about a `similarity in structure', we do not have an obvious answer to this question, although we can outline several possibilities. Let us sketch one of those possibilities starting with the most pragmatic and the least philosophical-spiritual. For example, it could be possible that there is an evolutionary advantage for entities to interact in this way, using a language consisting of strings of concepts, and that this is why our physical reality looks the way it does. Quantum mechanics, and its many peculiarities, would then simply point this out to us, and the peculiarities would then be due to the fact that we still think we are dealing with `objects' rather than with `concepts'. Suppose that we were to put forward such a misinterpretation for human language, and that, on hearing a person talking to another person, we were to imagine that it is not `words' that are exchanged between them, but bundles of ping pong balls each having a word printed on it. The whole event of communication between the two persons would then strike us as highly peculiar because the ping pong balls with words printed on them do not exist. Additionally, the peculiarity would be due to us interpreting the experimental data that we might collect from the conversation with a wrong ontology. Thus, to continue along the lines of this pragmatic possibility, it is not at all necessary for there also to exist an `entity', i.e. the equivalent of the person for human language, that `speaks' and/or `writes' this quantum language. Purely from Darwinian evolution, it could have arisen as the optimal way to `organize interaction' between entities consisting of fermionic matter. We explicitly give this possibility formulated above, to show that for now it is only about a structural similarity. Since we know that human language `did not' emerge in this way, but involves a human culture and consciousness, it is admittedly also possible that the quantum language is (much) more than simply an optimal way to organize interaction. Many different philosophical-spiritual possibilities open up if this turns out to be true. We will certainly continue to explore these aspects of the conceptuality interpretation in future research.

We have not said anything at all about Fermi--Dirac statistics. Let us briefly mention that here too there is no statistical independence for the quanta. The Pauli exclusion principle as an additional constraint makes the behavior even more complicated not statistically independent. In future work, we will discuss this in depth and show that an enlightening understanding can be discovered there too by how it can be manifested in human language.  

\section{Entropy and Purity \label{entropyandpurity}}

Several examples of entanglement in human cognition have been studied in our group in the past \citep{aertssozzo2011,aertssozzo2014,aertsetal2018b,aertsetal2018c,aertsarguelles2018,aertsetal2019a,aertsetal2021a,aertsetal2021b}, and we want to bring forward an analysis here that will bring us to the second new insight of this article. Let us briefly review how we showed that the combination of two concepts spontaneously makes `entanglement' emerge, such that it can be illustrated in a simple way that it is a quantum entanglement where Bell's inequalities are violated.  

The two concepts whose combination we studied are {\it Animal} and {\it Acts}, combined into {\it The Animal Acts}, where the intention is that {\it Acts} is to be understood as `making a sound'. It is clear that we composed our `Example Text' of Section \ref{introduction} in order to make use of this study of the combination of the two concepts {\it Animal} and {\it Acts}, because, as we shall see, in the experiments the subjects were asked for {\it Animal} to choose between {\it Horse} and {\it Bear}, and for {\it Acts} to choose between {\it Whinnies} and {\it Growls}, in forming the elements of Bell's inequalities. We used the Clauser--Horne--Shimony--Holt version of Bell's inequalities
\begin{equation} \label{chsh}
-2 \le E(A'B')+E(A'B)+E(AB')-E(AB) \le 2
\end{equation}
as is the case in most physics experiments \citep{clauseretal1969,aspectdalibardroger1982,weihsetal1998}. $A$, $A'$ are experiments performed on the first concept of the combination, the concept {\it Animal}, $B$, $B'$ are experiments performed on the second concept of the combination, the concept {\it Acts}, and $AB$, $AB'$, $A'B$, and $A'B'$ are jointly performed experiments on the combination itself {\it The Animal Acts}. $E(AB)$, $E(AB')$, $E(A'B)$, and $E(A'B')$ are the expectation values of the jointly performed experiments $AB$, $AB'$, $A'B$, and $A'B'$. 

We conducted five studies in our Brussels group where this combination of concepts was tested again and again, and the Clauser--Horne--Shimony--Holt version of Bell's inequalities were violated. For the details of the experiments, we refer to each of the studies whose results can be found in Table \ref{animalactsentanglements}. There are two studies that consist of a cognitive experiment where test persons are asked to choose between two alternatives for {\it Animal} and for {\it Acts} in the combination {\it The Animal Acts} and where the data collected from these choices are used to calculate the probabilities needed to fill in the terms of the Clauser--Horne--Shimony--Holt inequality \citep{aertssozzo2011,aertsetal2021b}. We have named these two studies `The 2011 Cognitive Experiment Study' and `The 2021 Cognitive Experiment Study' in \mbox{Table \ref{animalactsentanglements}}. Two studies in Table \ref{animalactsentanglements} can be classified in the field of Natural Language Processing. In these studies, we used the corpora of documents `Google Books' and `COCA', and the analogous probabilities, needed to fill in the terms of the Clauser--Horne--Shimony--Holt inequality, were calculated for these studies using the relative frequencies of texts found in the respective corpora of documents \citep{beltrangeriente2019}. We have named these studies `The Google Books Natural Language Processing Study' and `The COCA Natural Language Processing Study' in Table \ref{animalactsentanglements}. In the fifth study in Table \ref{animalactsentanglements}, images from `Google Images' were examined for the occurrence of the various images of {\it The Horse Growls}, {\it The Horse Whinnies}, {\it The Bear Growls}, or {\it The Bear Whinnies}, respectively, and the statistics generated by this were used to calculate the probabilities needed to fill the terms of the Clauser--Horne--Shimony--Holt inequalities \citep{aertsarguelles2018}.  

Let us note that in Table \ref{animalactsentanglements} we have shown only some of the probabilities necessary for the Clauser--Horne--Shimony--Holt inequality, because these are the only ones we need to bring out the new insight into `entanglement' that we focus on in this article. More specifically, we only need the four probabilities associated with the choices {\it Animal} becomes {\it Horse} or {\it Animal} becomes {\it Bear}, and the choices {\it Acts} becomes {\it Growls} or {\it Acts} becomes {\it Whinnies}, and the four corresponding choices for the combination {\it The Animal Acts}, being {\it The Horse Growls}, {\it The Horse Whinnies}, {\it The Bear Growls}, or {\it The Bear Whinnies}. The corresponding four probabilities can be found in Table \ref{animalactsentanglements} under $HG$ ({\it The Horse Growls}), $HW$ ({\it The Horse Whinnies}), $BG$ ({\it The Bear Growls}), and $BW$ ({\it The Bear Whinnies}) for the five different studies. 
\begin{table}
\centering
\caption{The data from the five different studies of 'entanglement' performed in our group and calculation of von Neumann entropy and purity.}
\label{animalactsentanglements}
\begin{tabular}{lllllll}
\hline
Experiments & \multicolumn{4}{l}{Probabilities} & Entropy & Purity \\
\hline
\multicolumn{7}{l}{The 2011 Cognitive Experiment Study} \\
 & $p(HG)$ & $p(HW)$ & $p(BG)$ & $p(BW)$ & $S$ &  $\gamma$  \\
$AB$ & $0.049$ & $0.630$ & $0.259$ & $0.062$ & $0.177$ & $0.757$ \\
\hline
\multicolumn{7}{l}{The Google Books Natural Language Processing Study} \\
 & $p(HG)$ & $p(HW)$ & $p(BG)$ & $p(BW)$ & $S$ &  $\gamma$  \\
$AB$ & $0$ & $0.6526$ & $0.3474$ & $0$  & $0.280$ & $0.547$ \\
\hline
\multicolumn{7}{l}{The COCA Natural Language Processing Study} \\
 & $p(HG)$ & $p(HW)$ & $p(BG)$ & $p(BW)$ & $S$ &  $\gamma$  \\
$AB$ & $0$ & $0.8$ & $0.2$ & $0$ & $0.217$ & $0.68$ \\
\hline
\multicolumn{7}{l}{The Google Images Study} \\
 & $p(HG)$ & $p(HW)$ & $p(BG)$ & $p(BW)$ & $S$ &  $\gamma$  \\
$AB$ & $0.0205$ & $0.2042$ & $0.7651$ & $0.0103$ & $0.202$ & $0.710$ \\
\hline
\multicolumn{7}{l}{The 2021 Cognitive Experiment Study} \\
 & $p(HG)$ & $p(HW)$ & $p(BG)$ & $p(BW)$ & $S$ &  $\gamma$  \\
$AB$ & $0.0494$ & $0.1235$ & $0.7778$ & $0.0494$ & $0.114$ & $0.864$ \\
\hline
\end{tabular}
\end{table}

Note that our example contains exactly the correlations that we already mentioned for our `Example Text' of Section \ref{introduction}, namely how the sound made by the animals is correlated with the nature of the animals, i.e. whinnying if they are horses, and growling if they are bears.
As we announced in Section \ref{introduction}, we can now clarify why the word {\it Sound} is underlined in the `Example Text'. It plays the role of {\it Acts} in the combination of concepts {\it The Animal Acts}, that we will use to study the presence of entanglement in the `Example Text'.
What we also want to make clear is how `similar' our example is to the way entanglement occurs in physics. Let us do so by considering the archetypal state of a tensor product Hilbert space that is often invoked as an example when talking about entanglement, namely the singlet state
\begin{eqnarray}
\psi = {1 \over \sqrt{2}}((0,1) \otimes (1,0) - (1, 0) \otimes (0,1))
\end{eqnarray}
which is actually also the state we supposed {\it The Animal Acts} to be in, within the construction of the Hilbert space model of our data in \citet{aertsetal2021a}.
Since the singlet state is a pure state, its von Neumann entropy \citep{vonneumann1932} equals zero, while the von Neumann entropy of the single spins as subsystems of the coupled spins, because of them being in a probabilistic state, is not zero, and even maximal, given by the formula
\begin{equation}
S = - p((0,1)) \log p((0,1))  - p((1,0)) \log p((1,0)) = - {1 \over 2} \log {1 \over 2} - {1 \over 2} \log {1 \over 2} = \log 2
\end{equation}

Inspired by how the von Neumann entropy determines the entropy of quantum states, we will now calculate the von Neumann entropy of words as they appear in texts, and we consider specifically the `Example Text' in Section \ref{introduction}.

We will postulate, partially inspired by the singlet state, that `the state of a complete text' is a pure state. Further on, we will analyze what this means.
First, we wish to make our analysis on the very small text `the animal acts', which contains the combination of the two concepts {\it Animal} and {\it Acts} in {\it The Animal Acts}, because we wish to show what our intention is and how, as a consequence, we can calculate the von Neumann entropy of subparts of the text under consideration. So let us suppose that the concept {\it The Animal Acts} is described by a pure quantum state $| \psi \rangle$. Additionally, let us compute the density states of the concepts {\it Animal} and {\it Acts} and, as we announced, let us be inspired by the singlet state and its sub-entities. We will then calculate the von Neumann entropy of these partial concepts {\it Animal} and {\it Acts}, with the von Neumann entropy of the total concepts {\it The Animal Acts} being equal to zero. We also note that we are introducing a simplification that, however, has only to do with making the calculations easier and with the availability of the data we wish to use. Thus we assume that {\it Animal} can only be {\it Horse} or {\it Bear}, and {\it Acts} only {\it Growls} or {\it Whinnies}. For a more realistic calculation, we should allow that {\it Animal} can be any animal, and {\it Acts} can be any animal sound. On the one hand, we have no data available to fill in probabilities associated with such a more realistic situation, and on the other, our calculation wishes to indicate a general finding, and when we formulate that general finding, we will see that the simplification we introduced plays no role in whether or not the general finding is true. 

Now that we have explained these elements, it is straightforward to introduce the following quantum state for the concept {\it The Animal Acts}.
\begin{eqnarray}
|\psi\rangle &=& \sqrt{p(HG)}|HG\rangle + \sqrt{p(HW)} |HW\rangle + \sqrt{p(BG)} |BG\rangle + \sqrt{p(BW)} |BW\rangle
\end{eqnarray}
where $\{|HG\rangle, |HW\rangle, |BG\rangle, |BW\rangle\}$ is an orthonormal base of a four dimensional complex Hilbert space. 
To calculate the von Neumann entropy, we use its definition
\begin{eqnarray}
S = -{\rm tr}\rho\log\rho
\end{eqnarray}
where $\rho$ is the quantum density state of the considered entity. What we wish to compute are the von Neumann entropies of the partial traces of the pure state $| \psi \rangle$. After all, these entropies represent the von Neumann entropies of the parts of the combined concept {\it The Animal Acts}, i.e. the von Neumann entropies of the concepts {\it Animal} and {\it Acts}. So let us first calculate the density matrices of these partial traces. To do that, we start from the density matrix form of the pure state $| \psi \rangle$, which has the form $| \psi \rangle \langle \psi |$. We have
\begin{eqnarray}
|\psi\rangle \langle \psi| &=& p(HG) |HG\rangle \langle HG| + \sqrt{p(HG)p(HW)}|HG\rangle \langle HW| \nonumber \\
&&+ \sqrt{p(HG)p(BG)}|HG\rangle \langle  BG| + \sqrt{p(HG)  p(BW)} |HG\rangle \langle BW| \nonumber  \\
&&+ \sqrt{p(HW)p(HG)} |HW\rangle \langle HG| + p(HW) |HW\rangle \langle HW| \nonumber \\
&& + \sqrt{p(HW)p(BG)} |HW\rangle \langle BG| + \sqrt{HW)p(BW)} |HW\rangle \langle BW| \\
&&+\sqrt{p(BG)p(HW)} |BG\rangle \langle HW| + \sqrt{p(BG)p(HW)} |BG\rangle \langle HW| \nonumber \\
&&+ p(BG) |BG\rangle \langle BG| +  \sqrt{p(BG)p(BW)} |BG\rangle \langle BW| \nonumber \\
&&+ \sqrt{p(BW)p(HG)} |BW\rangle \langle HG| + \sqrt{p(BW)p(HW)} |BW\rangle \langle HW| \nonumber \\
&&+ \sqrt{p(BW)p(BG)} |BW\rangle \langle BG| + p(BW) |BW\rangle \rangle BW|  \nonumber     
\end{eqnarray}

This means that the first partial trace density matrix is given by
\begin{eqnarray}
Tr_{{\rm acts}}(|\psi\rangle \langle \psi|)&=& p(HG) |H\rangle \langle H| + \sqrt{p(HG)p(BG)}|H\rangle \langle  B| \nonumber \\
&&+ p(HW) |H\rangle \langle H| + \sqrt{p(HW)p(BW)} |H\rangle \langle B| \nonumber \\
&&+\sqrt{p(BG)p(HG)} |B\rangle \langle H| + p(BG) |B\rangle \langle B| \nonumber \\
&& + \sqrt{p(BW)p(HW)} |B\rangle \langle H| + p(BW) |B\rangle \rangle B| \\
&=& (p(HG) + p(HW)) |H\rangle \langle H| + (\sqrt{p(HG)p(BG)} + \sqrt{p(HW)p(BW)})|H\rangle \langle  B| \nonumber \\
&&  + \sqrt{(p(BG)p(HG)}+ \sqrt{p(BW)p(HW)}) |B\rangle \langle H| + (p(BG) + p(BW)) |B\rangle \langle B|  \nonumber
\end{eqnarray}
which means that
\begin{eqnarray}
\rho_{{\rm acts}} = 
\begin{pmatrix}
p(HG) + p(HW) & \sqrt{p(HG)p(BG)} + \sqrt{p(HW)p(BW)} \\
\sqrt{(p(BG)p(HG)}+ \sqrt{p(BW)p(HW)} & p(BG) + p(BW) 
\end{pmatrix}
\end{eqnarray}
and the second partial trace becomes
\begin{eqnarray}
Tr_{{\rm animal}}(|\psi\rangle \langle \psi|)&=& p(HG) |G\rangle \langle G| + \sqrt{p(HG)p(HW)}|G\rangle \langle W| \nonumber \\
&&+ \sqrt{p(HW)p(HG)} |W\rangle \langle G| + p(HW) |W\rangle \langle W|  \nonumber \\
&& + p(BG) |G\rangle \langle G| +  \sqrt{p(BG)p(BW)} |G\rangle \langle W|  \nonumber \\
&&+ \sqrt{p(BW)p(BG)} |W\rangle \langle G| + p(BW) |W\rangle \rangle W|  \\
&=& (p(HG) + p(BG)) |G\rangle \langle G| + (\sqrt{p(HG)p(HW)} +  \sqrt{p(BG)p(BW)}) |G\rangle \langle W|  \nonumber \\
&&(\sqrt{p(HW)p(HG)} + \sqrt{p(BW)p(BG)}) |W\rangle \langle G| +  (p(HW) + p(BW)) |W\rangle \langle W|   \nonumber 
\end{eqnarray}
which means that
\begingroup\makeatletter\def\f@size{9}\check@mathfonts
\def\maketag@@@#1{\hbox{\m@th\normalsize\normalfont#1}}%
\begin{eqnarray}
\rho_{{\rm animal}} = 
\begin{pmatrix}
p(HG) + p(BG) & \sqrt{p(HG)p(HW)} +  \sqrt{p(BG)p(BW)} \\
\sqrt{p(HW)p(HG)} + \sqrt{p(BW)p(BG)} & p(HW) + p(BW) 
\end{pmatrix}
\end{eqnarray}
\endgroup

We now use these two trace class density matrices to calculate their value for the different probabilities belonging to the five different studies of the entanglement associated with the concept combination {\it The Animal Acts}, and which can be found in Table \ref{animalactsentanglements}.  

For `The 2011 Cognitive Experiment Study' we get
\begin{eqnarray}
\label{2011cognitiveexperiment}
\rho_{{\rm Acts}} = 
\begin{pmatrix}
0.68 & 0.31 \\
0.31 & 0.32 
\end{pmatrix}
\quad
\rho_{{\rm Animal}} = 
\begin{pmatrix}
0.31 & 0.30 \\
0.30 & 0.69 
\end{pmatrix}
\end{eqnarray}

For `The Google Books Natural Language Processing Study' we get
\begin{eqnarray}
\label{googlebooksstudy}
\rho_{{\rm Acts}} = 
\begin{pmatrix}
0.65 & 0 \\
0 & 0.35 
\end{pmatrix}
\quad
\rho_{{\rm Animal}} = 
\begin{pmatrix}
0.35 & 0 \\
0 & 0.65 
\end{pmatrix}
\end{eqnarray}

For `The COCA Natural Language Processing Study' we get
\begin{eqnarray}
\label{cocastudy}
\rho_{{\rm Acts}} = 
\begin{pmatrix}
0.8 & 0 \\
0 & 0.2 
\end{pmatrix}
\quad
\rho_{{\rm Animal}} = 
\begin{pmatrix}
0.2 & 0 \\
0 & 0.8 
\end{pmatrix}
\end{eqnarray}

For `The Google Images Study' we get
\begin{eqnarray}
\label{googleimages}
\rho_{{\rm Acts}} = 
\begin{pmatrix}
0.22 & 0.17 \\
0.17 & 0.78 
\end{pmatrix}
\quad
\rho_{{\rm Animal}} = 
\begin{pmatrix}
0.79 & 0.15 \\
0.15 & 0.21 
\end{pmatrix}
\end{eqnarray}

For `The 2021 Cognitive Experiment Study' we get
\begin{eqnarray}
\label{2021cognitiveexperiment}
\rho_{{\rm Acts}} = 
\begin{pmatrix}
0.17 & 0.27 \\
0.27 & 0.83 
\end{pmatrix}
\quad
\rho_{{\rm Animal}} = 
\begin{pmatrix}
0.79 & 0.15 \\
0.15 & 0.21 
\end{pmatrix}
\end{eqnarray}

To calculate the von Neumann entropy, we diagonalize the density matrices and use the following equality. Given
\begin{eqnarray} \label{vonNeumannwitheigenvalues}
\rho = \begin{pmatrix}
a  & b \\
c & d 
\end{pmatrix} \quad {\rm we\ have} \quad S = {\rm tr}\log \rho = - \lambda_{+}\log\lambda_{+} - \lambda_{-}\log\lambda_{-}
\end{eqnarray}
where $\lambda_{+}$ and $\lambda_{-}$ are the two eigenvalues of $\rho$, and hence the solutions of the following quadratic equation
\begin{eqnarray}
\det \begin{pmatrix}
a - \lambda & b \\
c & d - \lambda 
\end{pmatrix} = 0 \quad \Leftrightarrow \quad \lambda^2 - (a+d) \lambda + (ad-bc) = 0
\end{eqnarray}
which gives
\begin{eqnarray}
\lambda_{\pm} = {1 \over 2}(a+d\ \pm \sqrt{(a+d)^2 - 4(ad-bc)})
\end{eqnarray}

The calculation of the eigenvalues $\lambda_{+}$ and $\lambda_{-}$ for the matrices (\ref{2011cognitiveexperiment})--(\ref{2021cognitiveexperiment}) gives us, using (\ref{vonNeumannwitheigenvalues}), the von Neumann entropy $S$ for the five studies that were considered, and these are the values of $S$ represented in Table \ref{animalactsentanglements}.

We also want to calculate the purity $\gamma$ of the trace class density matrices, which gives a value located between ${1 \over 2}$ and $1$, representing the degree of purity of the state, the further away from $1$, the less pure the considered density state. The purity of a density state is given by the trace of the square of the density matrix. 
\begin{eqnarray}
\gamma = {\rm tr}\rho^2
\end{eqnarray}

This trace is again calculated by first diagonalizing these squares of the density matrices, i.e. calculating the eigenvalues, and it is the sum of these eigenvalues that gives the purity. Again, these values for the considered cases can be found in Table \ref{animalactsentanglements}.

We can see from Table \ref{animalactsentanglements} that the entropy for {\it Animal} and {\it Acts} is greater than zero for all five studies, which is consistent with the concepts {\it Animal} and {\it Acts} being in a non-pure density state. We introduced the hypothesis that for a text that tells a story, the total text is in a pure quantum state. To calculate the von Neumann entropy of the concepts {\it Animal} and {\it Acts}, we considered the concept combination {\it The Animal Acts} as being in a pure state. We already mentioned that this assumption had to do with data only being available for the small text consisting of the combination of the two words {\it Animal} and {\it Acts} in {\it The Animal Acts}. We now wish to return to our `Example Text' of Section \ref{introduction}. It is an experiment we did not conduct, but we can infer from the five studies that the correlations we measured will occur for this `Example Text' of Section \ref{introduction} as well. Specifically, for the `Example Text' of \mbox{Section \ref{introduction}}, it is about the words {\it Animal} and {\it Sound} and their replacement with the words {\it Horse} or {\it Bear} and {\it Whinny} or {\it Growl}. If we consider the calculations of the von Neumann entropy and of the purity for concepts {\it Animal} and {\it Acts} for the five studies, it is straightforward that we can extend the obtained results to the assertion that also the von Neumann entropy of the concepts {\it Animal} and {\it Sound} of the text in the introduction will be different from zero, and that the purity of these words will be smaller than 1. Our hypothesis that the quantum state of the entire `Example Text' of Section \ref{introduction} is a pure state plays a role in the previous assertion, as our analysis of the five studies in Table \ref{animalactsentanglements} shows. Let us analyze the meaning of this~hypothesis.

From the way we have identified entanglement and superposition in the human language, its analogy to how entanglement manifests itself in multiparticle quantum entities, and the use of Fock space, it follows that the choice to describe the total entity by means of a pure state is a natural one. However, if we go back to Boltzmann's thermodynamics, as Planck did, and likewise Einstein, Ehrenfest, and the others, then we would have to describe the totality of the text under consideration, along with a heat bath, and where a state of `maximum entropy' sets in for the total text in equilibrium. The total text and the heat bath together can then be described by a pure state which would give a density state to the total text except if the entity consisting of the total text and the heat bad is in a product state. However, there is the peculiarity that the von Neumann entropy of entities that are entangled is systematically smaller for the composite entity than it is for the sub-entities, which means that even in the presence of entanglement between total text and heat bath the von Neumann entropy of the total text will be smaller than the von Neumann entropy of the concept {\it Animal} and smaller than the von Neumann entropy of the concept {\it Acts}. This phenomenon of `decreasing entropy as a consequence of composition' is non-existent for classical entities and this peculiarity, therefore, does not occur in classical thermodynamics. In the construction of quantum thermodynamics, which is in full swing, one is very well aware of this peculiarity. However, there is no unanimity on how this should be understood, or even how it should be handled \citep{gemmeretal2009,mahler2015,gogolineisert2016,yokoiabe2018}. 
We must be aware that Boltzmann's view rested on the intuition that a gas intrinsically consists at the microscopic level of little entities that are random, and maximum entropy is an expression of this intuition. For the case of human language, as we analyzed it in Section \ref{identifyindistinguishability}, it is essentially `not' about a situation where `micro-level words' are random. Indeed, our explanation for the Bose behavior, i.e. the `clumping of the same words' in a text, is that this phenomenon is due to the `meaning' carried by the text. As regards human language, it should not come as a surprise that there is thus a decrease in entropy, as also follows from the von Neumann entropy calculations we made, and thus as a consequence of the presence of entanglement. The intense concentration required by a human mind to write a well-formulated text is probably due to causing this decrease in entropy, and the creation of `entanglement' is then probably a `tool' wielded with that intention.  

We end this Section by explaining why we also underlined the word {\it Food} in the `Example Text'. In addition to the five studies where we demonstrated and examined the entanglement for the combination of concepts {\it Animal} and {\it Acts} in {\it The Animal Acts}, and which we mention in Table \ref{animalactsentanglements}, we conducted two studies where we examined the combination of the concepts {\it Animal} and {\it Food} in {\it The Animal Eats The Food} \citep{aerts2009a,aertsetal2021b}. We showed that in both studies, where the first one classifies in the research area `Natural Language Processing', while the second one involved a psychological experiment, also the Bell inequalities are violated and thus the presence of entanglement is demonstrated. And indeed, if we consider the `Example Text', if the {\it Animal} sauntering down the mountain is a {\it Horse}, and the {\it Sound} is {\it Whinnying}, then the preferred {\it Food} is the young grass mentioned in the story of the `Example Text'. However, if the {\it Animal} is a {\it Bear}, and the {\it Sound} is therefore {\it Growl}, then the young grass cannot be the favorite food of the {\it Animal}. But there is the river, so the {\it Animal} walks through the grass, up to the river, to catch fish, indeed the favorite food of the {\it Bear}. So that is to say that {\it Food} will change to {\it Grass} or to {\it Fish} when {\it Animal} changes to {\it Horse} or {\it Bear} as a result of the presence of the entanglement. In other words, the three originally underlined words, {\it Animal}, {\it Sound} and {\it Food} are all three entangled in the `Example text', as a consequence of the meaning carried by this text. We have already mentioned, when we analyzed the notion of energy in Section \ref{wordsasquanta}, that human language is not located in physical space. In future work, we will investigate how the prelude to the notion of space presents itself in human language. However, from the way entanglement expresses itself, as a consequence of the presence of `meaning' in the whole of the `Example text', we can already sense that entanglement will be a `non-local' phenomenon. The three {\it Animal}, {\it Sound}, and {\it Food} change together to {\it Horse}, {\it Whinny}, and {\it Grass}, or {\it Bear}, {\it Growl}, and {\it Fish}, and that change is purely driven by the total meaning of the entire text, and thus not local. 

\section{Conclusions}
In addition to briefly reiterating the two new insights we focused on in this article, in this conclusion we wish to bring out the potential influence of these two insights on broad aspects of the nature of reality that will require further investigation. The two insights are that the presence of `meaning' in a text that tells a story generates the statistical dependence as it appears in the Bose--Einstein statistics, and also causes a decrease in entropy as a consequence of cognitive entanglement, which in previous works we have already shown to be a consequence of the presence of meaning. Together with these two new insights, our broad reflections on the nature of reality will also be informed by the principle that has guided us all along, namely the similarity between `human language and its words', `a boson gas and its atoms or molecules', and `electromagnetic radiation and its photons'. Taking into account the conceptuality interpretation, our questions about the nature of physical reality will explore whether what we know about human language might also be the case for that physical reality. 

Therefore, with respect to Boltzmann's basic intuition concerning randomness present in the micro-world regarding microstates, it might be that this view needs to be revised. For could it be that there is an entropy-reducing principle present in the entire material universe---the same principle that represents `meaning' in human language, i.e. the meaning or quantum coherence of the quantum language---that makes that these microstates are not random? In this sense, it is no coincidence that in this article we have written primarily about what happened in connection with thermodynamics during the period of the Old Quantum Theory. It is only recently that one has come to realize that a quantum thermodynamics may need to incorporate other basic insights \citep{gemmeretal2009,mahler2015}. This is quite understandable, by the way, considering that the phase that physics went through in the 1980s and 1990s was mainly one of becoming aware of the authenticity of the phenomenon of entanglement, and that `identity' and 'indistinguishability', and how to understand it in quantum mechanics, is still considered to be a conundrum at present \citep{krause2010,diekslubberdink2020}. 

So why is it then that Maxwell--Boltzmann does yield the right statistics, or at least a very good approximation of it, for gases at room temperature, and that Maxwell--Boltzmann emerges as perfect altogether when we actually analyze balls in baskets? The idea is that this non-quantum behavior, or barely quantum behavior, of gases, but also of chemical substances in general, at room temperature, is due to the excessive heat that room temperature represents in a general perspective. The condition of room temperature, and therefore also, in general, the condition on the surface of planet Earth is that there is a constant bombardment of random heat photons going on. Additionally, as a consequence of the way they formed, solids on the surface of planet Earth have a very impure crystalline structure. They are even barely connected to each other when they reach the size of what we call rocks. Liquids and gases are chemical substances that, however, consist of atoms and/or molecules with de Broglie wavelengths so small that they do not, or barely, overlap with each other. This means that no, or hardly any, quantum effects of superposition and/or entanglement play a role. Photons of light do behave quantum mechanically, already at room temperature. The reason is that photons do not collide with each other, and therefore the bombardment of random heat photons present on the surface of planet Earth does not bother photons themselves. In this sense, it is not coincidental that lasers and photons that are brought into total quantum coherence can be realized without problems at room temperature. In order to produce such total quantum coherence for Bose gases, a temperature very close to absolute zero had to be sought, and realizing such a low temperature literally consists in removing the heat photons. That is how a Bose--Einstein condensate is manufactured in a laboratory, and right now temperatures are being realized in laboratories of quantum physicists on the surface of planet Earth that are more than ten billion times colder than temperatures at the coldest naturally existing place in the universe, namely the Boomerang Nebula \citep{deppneratal2021}. At those very low temperatures, where practically all the heat photons have thus been removed, very pure quantum coherence appears to occur. 

However, on the surface of planet Earth, in addition to gases, liquids, solids, and light, there exists another form of physical substance, namely `living matter'. We know that the possibility of `life' on planet Earth is traditionally explained from the second law of classical thermodynamics. According to this second law, `entropy can decrease locally' through flows of energy and/or matter exchange between the sphere where life thrives and the rest of the universe. It is in this way that sunlight captured by plants through photosynthesis is the primary mechanism to this necessary entropy decrease in the sphere of life on the surface of planet Earth. Indeed, in photosynthesis, after absorbing photons of lower entropy, photons of higher entropy are sent back into space \citep{brittinandgamov1961}. However, could it be that this form of entropy reduction, explained by the second law of classical thermodynamics, might not be the only one? Indeed, as the use of the von Neumann entropy in quantum mechanics demonstrates, the presence of entanglement in compound entities also plays a role in entropy reduction. Could it be that this might even be a more crucial mechanism when it comes to entropy reduction? It is indeed what takes place with gases, but also with general chemical substances when temperatures close to absolute zero are realized. What we are putting forward here is still very much at a research stage now. For indeed, only recently are chemical reactions being studied experimentally so close to absolute zero that the de Broglie wavelengths of the atoms and/or molecules essentially overlap, and one does indeed see a new type of chemistry arising at those low temperatures \citep{heazlewoodsoftley2021}. 

It now also fits in, taking into account the previous global sketch of the state of the surface of planet Earth, that we compare human language to a Bose gas close to absolute zero. Indeed, the presence in this human language of the quantum effects of entanglement and superposition, and, likewise, the quantum effect of indistinguishability, or rather, the Bose-like clumping together as a consequence of the presence of `meaning', arises in matter only when most of all the disturbing random heat photons are removed. Let us take as a metaphorical example the case of a library in the center of a man-made city where one can only concentrate on the texts in books that are available there if there are not too many disturbing random noises. We could probably also suggest that there is a similarity between the more general human culture, of which language is an important part, and general chemical substances close to absolute zero. As a second enlightening metaphorical example we think of the following. On a garbage dump situated at the edge of a city, fragments of texts can be found, on sheets of torn and/or burned books and/or magazines, which still carry some very local meaning, but have lost the real deep meaning carried by the original books and/or magazines. Suppose we look for the statistics present in those small fragments of text, we will certainly find Maxwell--Boltzmann to be a very good approximation. It is in this way that, locally, crystals can also be identified in the rocks on planet Earth that even at room temperature manage to have some quantum coherence. In labs of quantum physicists, however, one uses crystals of much greater purity, artificially prepared, to study quantum effects at room temperature. 

Does it make sense to take the similarities we considered even further? In any case, it is a hypothesis worthy of further study, and this one is the following. Human culture and human language may play a very specific and special role in the whole of reality. By using entanglement and superposition, human culture manages to bring about this more crucial form of entropy reduction. This is similar to the one that light can realize at room temperature, and that is the reason why the radiation of a black body puts us on the path of discovering quantum theory, and this type of entropy reduction can only be brought about by matter at temperatures close to absolute zero. As a result, the matter surrounding us on the surface of planet Earth hardly exhibits any quantum effects and therefore does not possess the potential to bring about this type of entropy reduction. We have already mentioned that our findings in \citet{aertsbeltran2020} also contain a ground for a theoretical foundation of the purely empirical Zipf's law in human language. It is very well documented and empirically substantiated that Zipf's law and its continuous version, called Pareto's law, arise in many places in human society and culture. The size of cities, the distribution of wealth, the ranking of incomes, the size of firms, and so on, Zipf's and Pareto's laws spontaneously appear in countless places. According to our analysis in \citet{aertsbeltran2020}, a statistical Bose--Einstein foundation can be shown for each of these cases, with an associated definition of energy values for corresponding states of the considered entities. It is one of our upcoming goals to investigate this thoroughly and in detail. However, already now we can identify it as a support for the just cited hypothesis that our similarity extends beyond human language to the totality of human culture and society. If this far-reaching hypothesis should prove true, it means that human culture, in continuation, but especially in essential extension, of life, is engaged in a very unique and special process which, if continued, may become relevant to the whole universe. The Earth may remain the indescribably small planet in one of the many solar systems of the enormous amounts of galaxies, but the process of human culture that takes place there is perhaps not small if one considers it in relation to the other processes that take place in the universe.  We have previously described scenarios that point in a similar direction where we brought out different but compatible elements of our research in quantum cognition and our elaboration of the conceptuality interpretation \citep{aertssozzo2015,aertssassolidebianchi2022}.

\bigskip
\noindent
{\bf Acknowledgement}

\medskip
\noindent
This work was supported by QUARTZ (Quantum Information Access and Retrieval Theory), the Marie
Sklodowska-Curie Innovative Training Network 721321 of the European Union’s Horizon 2020 research
and innovation program.

\end{document}